\documentclass{amsart}
\usepackage{mathptmx,fullpage,amsmath,amssymb,amsfonts,wrapfig,graphicx,tikz,caption,float,color,enumerate,wrapfig}
\usepackage{graphicx,subcaption}
\usetikzlibrary{braids,positioning}
\tikzset{main node/.style={circle,fill=blue!20,draw,minimum size=1cm,inner sep=0pt},}

\definecolor{mycolor}{rgb}{0, 0, 0.6}
\usepackage[colorlinks,urlcolor=blue,
    citecolor=mycolor,linkcolor=mycolor]
    {hyperref}

\newtheorem{thm}{Theorem}
\newtheorem*{reid}{Reidemeister's Theorem}
\newtheorem*{karp}{Karp's Theorem}
\newtheorem{lem}[thm]{Lemma}
\newtheorem{cor}[thm]{Corollary}

\newcommand{\p}{\mathsf{P}}
\newcommand{\NP}{\mathsf{NP}}
\newcommand{\RR}{\mathbb{R}}
\newcommand{\ZZ}{\mathbb{Z}}
\newcommand{\lk}{lk}
\newcommand{\sign}{sign}

\newcommand{\defeq}{\stackrel{\text{def}}{=}}

\begin{document}

\title{An elementary proof that linking problems are hard}

\author{Shannon Cheng, Anna Chlopecki, Saarah Nazar and Eric Samperton}
\address{Departments of Mathematics and Computer Science, Purdue Quantum Science \& Engineering Institute (PQSEI), 150 N. University St., West Lafayette, IN 47907}
\thanks{All authors supported in part by NSF CCF 2330130.  We thank the Purdue Experimental Math Lab---especially its inaugural director Thomas Sinclair---for bringing us together.
We also thank our anonymous referee for many helpful suggestions.}

\date{November 24, 2026}

\begin{abstract}
We give a new, elementary proof of what we believe is the simplest known example of a ``natural'' problem in computational 3-dimensional topology that is $\mathsf{NP}$-hard---namely,
the \emph{Trivial Sublink Problem}: given a diagram $L$ of a link in $S^3$ and a positive integer $k$, decide if $L$ contains a $k$ component sublink that is trivial.
This problem was previously shown to be $\mathsf{NP}$-hard in independent works of Koenig-Tsvietkova and de Mesmay-Rieck-Sedgwick-Tancer, both of which used reductions from $\mathsf{3SAT}$.
The reduction we describe instead starts with the Independent Set Problem, and allows us to avoid the use of Brunnian links such as the Borromean rings.
On the technical level, this entails a new conceptual insight: the Trivial Sublink Problem is hard entirely due to mod 2 pairwise linking, with no need for integral or higher order linking.
On the pedagogical level, the reduction we describe is entirely elementary, and thus suitable for introducing undergraduates and non-experts to complexity-theoretic low-dimensional topology.
To drive this point home, in this work we assume no familiarity with low-dimensional topology, and---other than Reidemeister's Theorem and Karp's result that the Clique Problem is $\mathsf{NP}$-hard---we provide more-or-less complete definitions and proofs.
We have also constructed a web app that accompanies this work and allows a user to visualize the new reduction interactively.
\end{abstract}

\maketitle

\begin{figure}[h]
\includegraphics[width=0.75\textwidth]{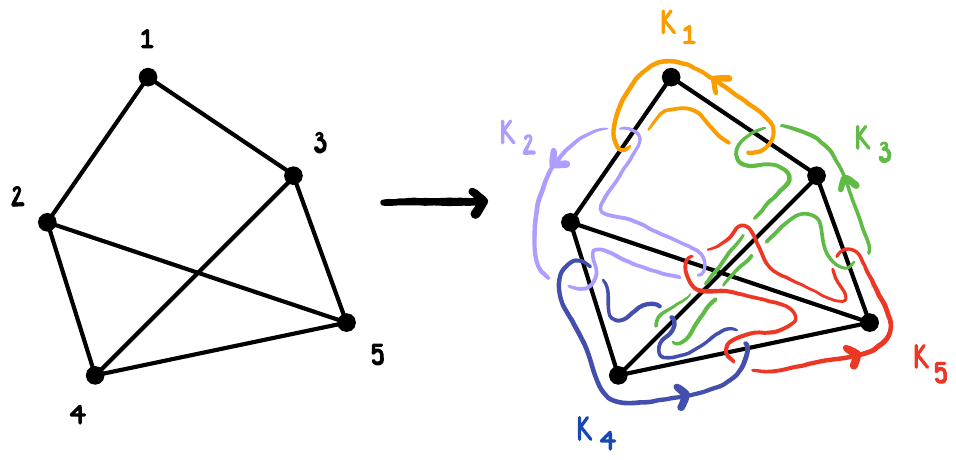}
    \caption{Tancer's ``octopus" reduction from Independent Set to Trivial Sublink.}
    \label{fig-martin2}
\end{figure}

\section{Introduction}
\label{sec-intro}
\subsection{Motivations and caveats}
\label{ss-motivations}
Knots have apparently fascinated humans longer than the written word has existed.
A common subtext in this fascination is the ``intricacy" or ``complexity" of knots, as illustrated by the Dharmic use of the Endless Knot for at least four thousand years, the mythical Gordian Knot of classical antiquity, or the proliferation of Celtic knots in early Insular art.
The more recent advent of computers---and concomitant developments in the language of theoretical computer science---has allowed humans to make mathematically precise statements about the ``complexity of knots."
In particular, it is now generally understood that there are many different precise problems one can associate to knots, each of whose complexities can be understood as different ways that one can measure the ``complexity of knots."

With this in mind, our goal in this work is to give the simplest known formulation and proof of the idea that ``knots are complex."
By doing so, we hope to encourage more people to think about complexity-theoretic knot theory (especially the more important problems that we will identify along the way).\footnote{For the sake of space, however, we will not provide much more \emph{motivation} for knot theory beyond the couple of paragraphs above.
Suffice it to say that there are plenty of good reasons to study knots, both from the purely mathematical side and for applications (e.g., DNA topology or topological quantum physics).
Our ideal reader is someone who is already curious about computational aspects of knot theory, but maybe does not know where to begin.
}

Before really diving in, we should fess up to a few things.
First, our precise results will not technically be about ``knots," but rather ``links," which are like knots except that they are allowed to have several different connected components (think several different ropes knotted together; each individual rope is a component).
Second, the specific computational problem for links that we will study is perhaps a bit contrived; however, we believe it is interesting enough to take seriously as an illustration of the computational richness of low-dimensional topology, while simultaneously being easy enough to explain in an elementary way.

Third, our main result---see Theorem \ref{thm-main} below---has already been proved before.
Figure \ref{fig-martin2} shows the idea of one prior proof; we discuss it in some detail in Subsection \ref{ss-main}, but for experts, it is probably enough to understand everything just by looking at it now.
In the details, we will provide a \emph{new} proof using \emph{braids}, which we believe is useful pedagogically as it allows us to unpack everything so explicitly that it can be implemented easily in Python; see the URL provided in Section \ref{ss-main}.
Also, the explicitness of our braid strategy makes us suspect it could be useful in other settings.

Our proof is similar in spirit to the strategy in Figure \ref{fig-martin2}.
In technical language (that we will explain in due course), the key is to convert a graph $G$ encoded via an adjacency matrix $A$ into a link diagram $L_G$ where each component is an unknot, and the (mod 2) linking matrix of $L_G$ equals $A$ (for spoilers, the reader can look ahead all the way to the final Figure \ref{fig-example}).
We note that our way of achieving this via braid words does not typically output the same link diagram (or even link type) as Tancer's ``octopus" reduction in Figure \ref{fig-martin2}.

Finally, the proof strategy in Figure \ref{fig-martin2} was first explained to one of us by Yo'av Rieck in a five minute lightning talk given during the 2023 Computational Problems in Low-dimensional Topology III workshop at Rutgers University-Newark.
During that talk, Rieck credited the main idea to Martin Tancer, and it has since appeared in print in de Mesmay's work \cite{demesmay}..
We are grateful to both Rieck and Tancer for their generosity in sharing the idea, as well as some helpful comments from Tancer on the first draft of this work.
We also thank the organizers of the workshop Eric Sedgwick, Anastasiia Tsvietkova, and Mehdi Yazdi.

\subsection*{Outline}
In the next two subsections \S\ref{ss-complexity} and \S\ref{ss-knots}, we review the basics of computational complexity theory and computational knot theory, respectively.  In \S\ref{ss-main} we formally introduce our main result Theorem \ref{thm-main} and discuss some simple consequences.

Section \ref{sec-proof} constitutes the remainder of the paper and the proof of Theorem \ref{thm-main}.  Subsection \S\ref{ss-graphs} reviews the basics of graphs, cliques, independent sets and Karp's theorem that the clique problem is $\NP$-hard.  Subsection \S\ref{ss-linking} introduces the key notion of linking matrix of a link.  \S\ref{ss-braids} and \S\ref{ss-pure} introduce useful concepts from the theory of braids, and \S\ref{ss-splitting} introduces a simple criterion for determining when two components of a link can be ``split" away from one another.  Finally, we complete the proof of the main theorem in \S\ref{ss-reduction}.

\subsection{Complexity theory 101}
\label{ss-complexity}
Before diving into computational knot theory, we should lay down some of the basics of complexity theory.
We won't bother with getting into the rigorous foundations (Turing machines, \emph{etc.}), so the following definitions should be understood as informal, even though they can certainly be made formal---consult with any of the numerous textbooks on complexity theory for more info (our favorite is \cite{arora2009computational}).
Admittedly, this section is quite terse---we recommend reading it quickly at first, and referring back to it as you read later sections on an as-needed basis.

First, as a convenience, in complexity theory we usually concern ourselves with \emph{decision problems}, which are simply ``yes/no questions" whose input is an arbitrary bit string:
\[\begin{aligned}
d: \{0,1\}^* \defeq \bigsqcup_{k \ge 1} \{0,1\}^k &\to \{0 , 1\} \\
x &\mapsto d(x).
\end{aligned}\]
As a matter of interpretation, we think of the function $d$ as a family of questions, $x$ as a specific question from that family, $d(x)=0$ as a ``No" answer to the specific question $x$, and $d(x)=1$ as a ``Yes" answer.
Moreover, we should note that we do not typically care about bit strings \emph{per se}.
Rather, the bit string $x$ is usually understood as encoding some other type of mathematical object, in the same way in which, say, a graph described via an adjacency matrix inside the memory of a computer is ultimately encoded as some bit string describing transistors' states.
If $x \in \{0,1\}^k$, then $x$ is of course a bit string of length $k$, and we may write this in notation as $|x|=k$.
In practice, we always work with higher-level---but still finite and discrete!---data structures (such as graphs, knot diagrams, \emph{etc.}) where we do not need to bother with the specifics of how those data structures are encoded specifically as bit strings, so long as we understand how the size of the bit string is controlled by the size of the relevant data structure.
We will see several examples of all this in due course.

We say a decision problem $d$ is \emph{recursively enumerable} if there exists an algorithm with the following two properties:
\begin{enumerate}[(i)]
\item For all $x \in \{0,1\}^*$ with $d(x)=1$, the algorithm correctly outputs the fact $d(x)=1$ in a finite amount of time when provided $x$ as input.
\item For all $x \in \{0,1\}^*$ with $d(x)=0$, the algorithm loops infinitely, never reporting any output, when provided $x$ as input.
\end{enumerate}
The set of all recursively enumerable decision problems is denoted $\mathsf{RE}$.
This is our first example of a complexity class.

Dually, a decision problem $d$ is \emph{co-recursively enumerable} if there exists an algorithm with the following two properties:
\begin{enumerate}[(i)]
\item For all $x \in \{0,1\}^*$ with $d(x)=0$, the algorithm correctly outputs the fact $d(x)=0$ in a finite amount of time when provided $x$ as input.
\item For all $x \in \{0,1\}^*$ with $d(x)=1$, the algorithm loops infinitely, never reporting any output, when provided $x$ as input.
\end{enumerate}
For a problem to be recursively enumerable, we do not require anything about the algorithm.
The set of all co-recursively enumerable decision problems is denoted $\mathsf{coRE}$.

A decision problem $d$ is \emph{computable} (or \emph{recursive}, or \emph{decidable}) if there exists an algorithm which, when given any bit string $x$ as input, correctly outputs the value $d(x)$ in a finite amount of time.
The set of all computable decision problems is denoted $\mathsf{R}$.
It is a fun exercise to check that $\mathsf{R} = \mathsf{RE}\cap\mathsf{coRE}$.

\begin{wrapfigure}{r}{0.3\textwidth}
\centering
\begin{tikzpicture}
  \node (RE) at (1,6) {$\mathsf{RE}$};
  \node (coRE) at (-1,6) {$\mathsf{coRE}$};
  \node (R) at (0,5) {$\mathsf{R}$};
  \node (ER) at (0,4) {$\mathsf{ER}$};
  \node (EXP) at (0,3) {$\mathsf{EXP}$};
  \node (PSPACE) at (0,2) {$\mathsf{PSPACE}$};
  \node (NP) at (-1,1) {$\mathsf{NP}$};
  \node (coNP) at (1,1) {$\mathsf{coNP}$};
  \node (P) at (0,0) {$\mathsf{P}$};
  \draw (P) -- (NP);
  \draw (P) -- (coNP);
  \draw (NP) -- (PSPACE);
  \draw (coNP) -- (PSPACE);
  \draw (PSPACE) -- (EXP);
  \draw (EXP) -- (ER);
  \draw (ER) -- (R);
  \draw (R) -- (RE);
  \draw (R) -- (coRE);
\end{tikzpicture}
\caption{Some of the basic complexity classes and their containment relations.}
\label{f:cc}
\end{wrapfigure}

\emph{Computability theory}, roughly speaking, is concerned with understanding which decisions problems belong to $\mathsf{R}$.
\emph{Complexity theory} asks for something more: once a problem is known to be computable, how easy or hard is it to \emph{actually} compute it?
To make this precise, we introduce complexity classes that are defined via various ``resource constraints."
Here are four important examples:

\begin{enumerate}
\item $\mathsf{EXP}$ (exponential time): a decision problem $d$ is in $\mathsf{EXP}$ if there exists a positive integer $b$ and an algorithm which, when given a bit string $x$ with $|x|=k$, correctly outputs $d(x)$ in time $O(2^{k^b})$.\footnote{Recall that given two functions $f$ and $g$ with domain containing an unbounded subset of $\mathbb{R}_{\ge 0}$, we write $f=O(g)$ (read ``$f$ is big O of $g$") to mean that there exist real constants $N, C>0$ such that for all $x>N$, $f(x)\le C\cdot g(x)$.}
\item $\p$ (polynomial time): a decision problem $d$ is in $\p$ if there exists a positive integer $b$ and an algorithm which, when given a bit string $x$ with $|x|=k$, correctly outputs $d(x)$ in time $O(k^b)$.
\item $\mathsf{PSPACE}$ (polynomial space): a decision problem $d$ is in $\mathsf{PSPACE}$ if there exists a positive integer $b$ and an algorithm which, when given a bit string $x$ with $|x|=k$, correctly outputs $d(x)$ and uses $O(k^b)$ space.
\item $\mathsf{ER}$ (elementary recursive): a decision problem $d$ is in $\mathsf{ER}$ if there exists a positive integer $b$ and an algorithm which, when given a bit string $x$ with $|x|=k$, correctly outputs $d(x)$ in time \[O\left(2^{2^{\cdots^{2^k}}}\right)\]
where the height of the exponential tower is $b$.
\end{enumerate}

We note that problems in $\p$ are generally understood to be the problems that admit ``efficient" algorithmic solutions.
As far as practical applications go, the goal of complexity theory can be over-simplified as follows: determine exactly which problems are in $\p$, and which are not.

When we define complexity classes, there is nothing stopping us from introducing ``unrealistic" resources.
This may seem silly, but in fact, doing this leads to one of the most important and infamous complexity classes: $\NP$.
Informally, a decision problem $d$ is in $\NP$ if for all $x$ with $d(x)=1$, there is a short and easy-to-check proof that $d(x)=1$.
Somewhat less informally, we need there to exist a ``short" bit string $w$---sometimes called a \emph{certificate}, or \emph{witness}---such that we can use $w$ to convince ourselves that $d(x)$ does in fact equal $1$ using an algorithm that runs in time polynomial in $|x|$.

More formally: a decision problem $d$ is in the complexity class $\NP$ (non-deterministic polynomial time) if there exist polynomials $p$ and $q$, and an algorithm, such that
\begin{enumerate}[(i)]
\item for all $x$ with $d(x)=1$, there exists a $w \in \{0,1\}^{p(|x|)}$ such that, when input $(x,w)$, the algorithm correctly outputs $d(x)=1$,
\item for all $x$ with $d(x)=0$ and $w \in \{0,1\}^{p(|x|)}$, the algorithm always outputs $d(x)=0$ when given $(x,w)$ as input,
\end{enumerate}
and, moreover, for all $(x,w)$ the algorithm runs in time $O(q(|x|))$.

In this definition of $\NP$, condition (i) is sometimes called the \emph{completeness} of the non-deterministic algorithm---the connotation being that for all $d(x)=1$, there does in fact exist a $w$ that can be used to certify that $d(x)=1$.
Likewise, condition (ii) is sometimes called the \emph{soundness} of the non-deterministic algorithm---if $d(x)=0$, the algorithm will never be tricked by any $w$ into incorrectly asserting that $d(x)=1$.

Similarly to the relation between $\mathsf{RE}$ and $\mathsf{coRE}$, we can define a ``dual" complexity class $\mathsf{co}\NP$ to $\NP$ by exchanging the roles of $1=$Yes and $0=$No as follows:
a decision problem $d$ is in the complexity class $\mathsf{co}\NP$ (non-deterministic polynomial time) if there exist polynomials $p$ and $q$, and an algorithm, such that
\begin{enumerate}[(i)]
\item for all $x$ with $d(x)=0$, there exists a $w \in \{0,1\}^{p(|x|)}$ such that, when input $(x,w)$, the algorithm correctly outputs $d(x)=0$,
\item for all $x$ with $d(x)=1$ and $w \in \{0,1\}^{p(|x|)}$, the algorithm always outputs $d(x)=1$ when given $(x,w)$ as input,
\end{enumerate}
and, moreover, for all $(x,w)$ the algorithm runs in time $O(q(|x|))$.

In contrast to the fact $R = \mathsf{RE} \cap \mathsf{coRE}$, we note that it is \emph{not} generally believed that $P = \NP\cap \mathsf{co}\NP$---but this is an open problem!

We note that saying a problem is in $\NP$ does not do much for ``realistic" algorithms to compute $d(x)$, since it does nothing to tell us \emph{how} to find a valid certificate when it exists---much less tell us whether or not one exists in the first place.
Nevertheless, $\NP$ plays a critical role in complexity theory.
In practice, it is often straightforward to understand whether or not a given problem $d$ is in $\NP$.
If it is, then it is sensible to start wondering whether or not $d$ is in $\p$; if it is not, then it is presumably unfeasible to expect to solve $d$ algorithmically.

However, this strategy for understanding which problems are in $\p$ runs into a cononudrum: there appear to be some problems in $\NP$ that are harder than others, and so putting a problem in $\NP$ turns out not to be a very strong indicator for whether or not the problem is in $\p$.

To make this precise, we must understand \emph{reductions}.
Note that, loosely speaking, putting a decision problem into some specific complexity class is a kind of ``upper bound" on the complexity of that problem.
Doing so says that we do not need more than so-and-so many resources to solve the problem algorithmically.
Reductions, on the other hand, allow us to put a ``lower bound" on a problem's complexity, by saying that the problem is at least as hard as some other problem.

There are in fact many different precise notions of reduction, but we will only need one: Karp reduction.

Let $d_0$ and $d_1$ be two decision problems.
A \emph{Karp reduction from $d_0$ to $d_1$} is an algorithm $r$ that converts each bit string $x$ into a new bit string $r(x)$ in such a way that $d_1(r(x)) = d_0(x)$ and in time polynomial in $|x|$.
In particular, if we had a polynomial time algorithm to solve $d_1$, then we would have one to solve $d_0$ too.

In practice, when we show that one problem $d_0$ admits a Karp reduction $r$ to another problem $d_1$, we do this in two parallel parts.
In one part---the ``completeness" portion of the reduction proof---we show that if $d_0(x)=1$, then $d_1(r(x))=1$.
In the other part---the ``soundness" portion of the reduction proof---we show that if $d_0(x)=0$, then $d_1(r(x))=0$.

Finally, we say a problem $d$ is \emph{$\NP$-hard} if for all problems $d_0 \in \NP$, there exists a Karp reduction from $d_0$ to $d$.
If $d$ is both in $\NP$ and $\NP$-hard, then we say that $d$ is \emph{$\NP$-complete}.

It is easy to argue check from the definitions that $\p\subseteq \NP$.
The famous $\p$ versus $\NP$ conjecture asks whether or not there exist problems in $\NP$ that are not in $\p$.
A stronger conjecture, called the exponential time hypothesis, posits that the best possible running time for an $\NP$-hard problem is exponential.

Examples will be forthcoming, but let's turn to knots.

\subsection{Computational knot theory 101}
\label{ss-knots}
For mathematicians, a \emph{knot} is a smooth embedding $K: S^1 \to \RR^3$, where $S^1$ is the unit circle and $\RR^3$ is 3-dimensional Euclidean space.\footnote{A map is \emph{smooth} (or $C^\infty$) if it has continuous derivatives of all order.
A \emph{smooth embedding} is a continuous map that is a homeomorphism onto its image and whose Jacobian has full rank at all points in its domain.
We could instead define everything using \emph{piece-wise linear} maps, and the two theories would be the same.
Since we will quickly move on to encoding knots and links via diagrams, we refer the reader to \cite{CrowellFox} for discussion of these matters.}
A \emph{link with $n$ components} is a smooth embedding
\[L: \bigsqcup_{i=1}^n S^1 \to \RR^3,\]
where $\bigsqcup_{i=1}^n S^1$ is the disjoint union of $n$ circles.
Intuitively, a knot is what we get by taking a single length of (arbitrarily thin) rope, tying a knot in it, and then gluing its two ends together.
An $n$ component link is what we get by doing this to $n$ different lengths of rope all at the same time, leading to ``several knots knotted together."
In this analogy, each individual rope is one of the components of the link.

Two links $L_0$ and $L_1$ are considered \emph{equivalent} if they have the same number of components $n$ and are \emph{(smoothly) ambient isotopic}, meaning there exists a smooth map $H: \RR^3 \times [0,1] \to \RR^3$ with the following three properties:
\begin{itemize}
\item for all $t \in [0,1]$, the map $x \mapsto H(x,t)$ is a diffeomorphism\footnote{A \emph{diffeomorphism} is a smooth, invertible map whose inverse is also smooth (understanding this is not important in this paper).} of $\RR^3$,
\item the map $x \mapsto H(x,0)$ is the identity map on $\RR^3$, and
\item for all $\theta \in  \bigsqcup_{i=1}^n S^1$, $H(L_0(\theta),1) = L_1(\theta)$.
\end{itemize}
It is a fun exercise to check that this yields an equivalence relation on the set of all links.
This notion of equivalence precisely captures the intuitive idea that two tangled up messes of circular ropes should be considered the same if we can wiggle and stretch and shrink one (without cutting it) until it ``looks like" the other one.
The fundamental topological problem of knot theory is, in some sense, to classify all possible equivalence classes of knots and links.

(One technical point: our definition of ``equivalent" here implies that we are really working with \emph{oriented} knots and links.  That is, we think of $S^1$ as being endowed with its standard orientation, and we keep track of it.  This does not affect our main result at all, but it is convenient for our proof.)

These definitions (knot, link, ambient isotopic, equivalent) are all well and good as far as rigorous mathematics is concerned, but are ill suited to computation.
Indeed, \emph{a priori}, a smooth embedding $K: S^1 \to \RR^3$ typically requires an uncountably infinite amount of data to encode.
Fortunately, when studying the topology of knots, we are ultimately more interested in understanding the equivalence classes of knots---not specific knots \emph{per se}---which gives us flexibility to combinatorialize our representations of knots.
For example, one can replace the assumption that a knot is a \emph{smooth} map $K: S^1 \to \RR^3$ with the assumption that it is \emph{piecewise-linear} map, which can always be isotoped (in the piecewise-linear sense) so that its ``kinks" meet at rational points $\mathbb{Q}^3 \subset \RR^3$.

The more common way to combinatorialize descriptions of links is to use 2-d pictures called ``diagrams", which combine a plane graph describing the ``shadow" of the link we get by projecting it onto a plane $\RR^2\subset\RR^3$ with the ``crossing information" that indicates which parts of this shadow come from pieces of the knot that are ``above" others.
We will do it, but warn the reader before doing so that it is a bit obnoxious to write down a rigorous mathematical definition for what is supposed to be a very simple idea.
Before reading the next paragraph, we encourage the reader to skim the figures throughout this document, which contain many examples of link diagrams (as well as ``braid words"---more to come on that!).

A \emph{link diagram} $D$ is a 4-regular, plane graph (that is, a planar graph together with a choice of (isotopy class of) embedding in the plane; for a review of the basic definitions from graph theory, see \S\ref{ss-graphs}) such that at every vertex, an opposite pair of  half-edges is distinguished by calling them the \emph{undercrossing half-edges}.
We call the other two half-edges the \emph{overcrossing half-edges} and we call vertices \emph{crossings}.
Note that the graph underlying a link diagram $D$ need not be simple, \emph{i.e.}\ there can be self-loops at the vertices, as well as multiple edges between two vertices.
Moreover, the graph is allowed to have connected components that do not contain any vertices, called \emph{loops}.

\begin{figure}
\centering
\begin{subfigure}[t]{0.3\textwidth}
\centering
\includegraphics[width=0.9\textwidth]{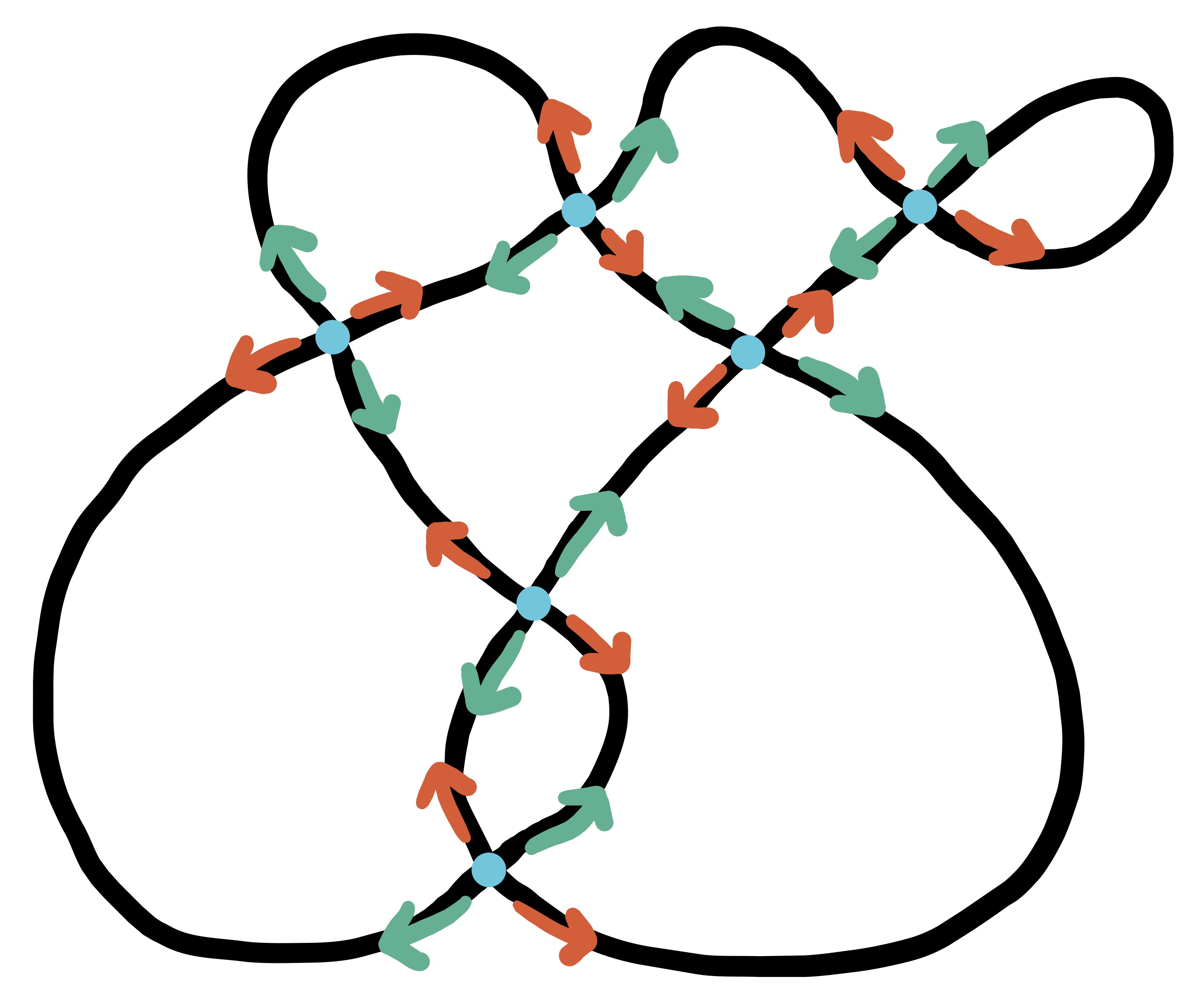}
\caption{The ``formal" data structure representing a diagram of the $5_2$ knot, with the undercrossing half-edges colored green, and the overcrossing ones colored red.}
\end{subfigure}
\qquad
\begin{subfigure}[t]{0.3\textwidth}
\centering
\includegraphics[width=0.9\textwidth]{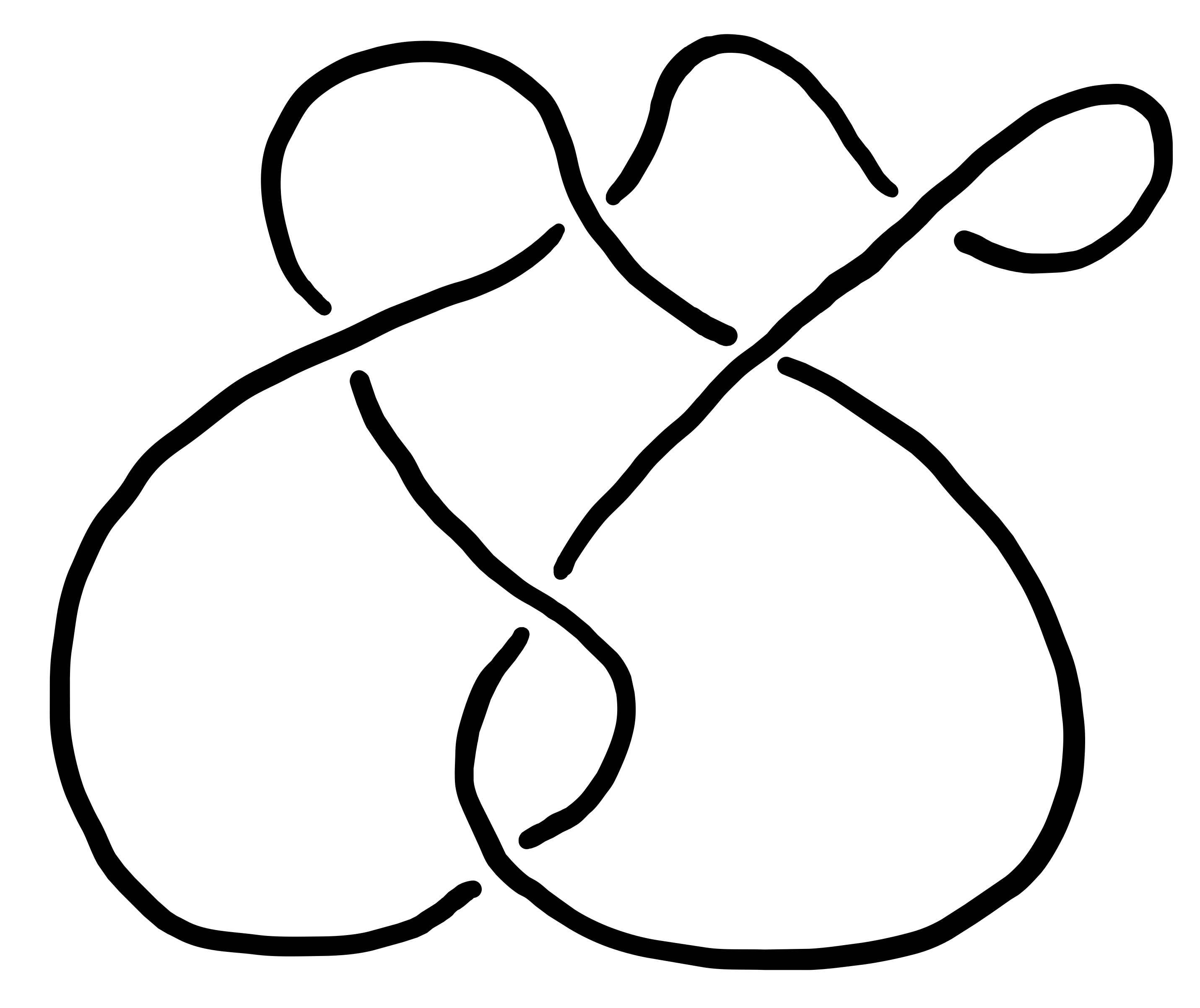}
\caption{The usual way we draw the diagram from (A).}
\end{subfigure}
\qquad
\begin{subfigure}[t]{0.3\textwidth}
\centering
\includegraphics[width=0.9\textwidth]{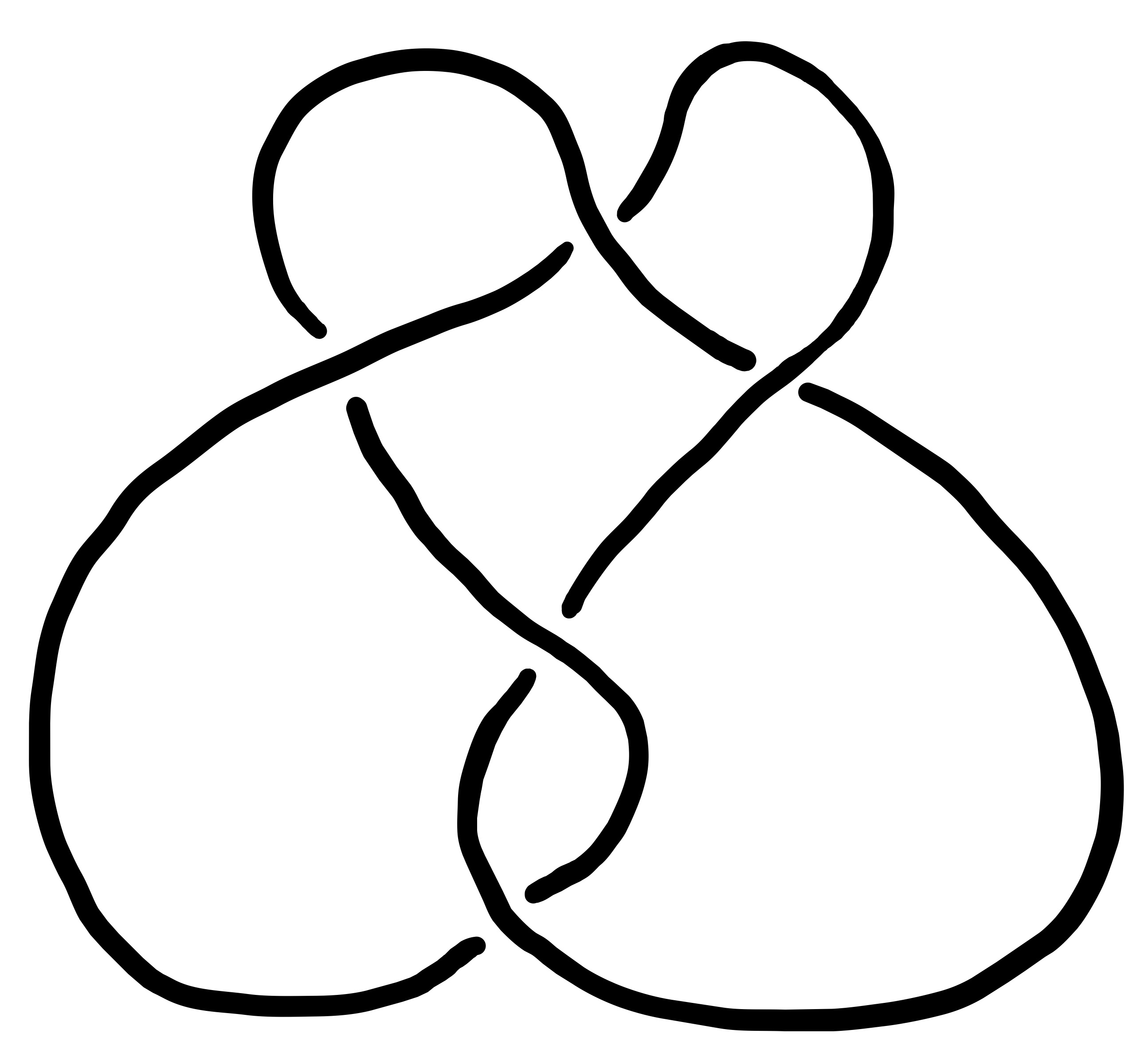}
\caption{A ``simpler" diagram of the same knot from (B) after applying an R1 move.}
\end{subfigure}
\label{fig:5_2_knot}
\caption{}
\end{figure}

Every link diagram determines a partition of its edges and loops into parts called \emph{components} as follows.
Each loop is defined to be in its own component.
If two edges meet at opposite sides of a vertex, then they are declared to be in the same component; this binary relation generates a partition on the edges whose parts are then the components.

An \emph{orientation} of a link diagram is a choice of loop orientations and edge orientations such that at each pair of opposite half-edges of each vertex, one of the half-edge's orientations points toward the vertex and the other's points away from the vertex.
Every link diagram can be oriented.

Given any $n$ component link $L$, there exists an oriented link diagram $D_L$ with $n$ components that---in a sense we make precise momentarily---``represents" $L$.
We sketch the basic idea in three steps.
First, apply an (arbitrarily small) ambient isotopy of $L$ so that it is in ``general position" with respect to the orthogonal projection map $\pi: \RR^3 \to \RR^2$, where $\RR^2$ is the $xy$ plane in $\RR^3$.
Second, thanks to this ``general position" condition, the projection $\pi(L)$ of $L$ onto $\RR^2$ is a 4-regular plane graph with loops.
Finally, we determine crossing information at the vertices by keeping track of which half edges come from points of $L$ with larger $z$ coordinates.

Conversely, given an oriented link diagram $D$, we can  build a link $L_D$ in such a way that $L_{D_L}$ is equivalent to $L$.
It is not too hard to work out by oneself how this should be done.
We call $L_D$ the link \emph{represented} by the diagram $D$.

The question of whether or not two links $L$ and $L'$ are equivalent can be entirely addressed using link diagrams representing them, a claim made precise by

\begin{figure}
\centering
\includegraphics[width=\textwidth]{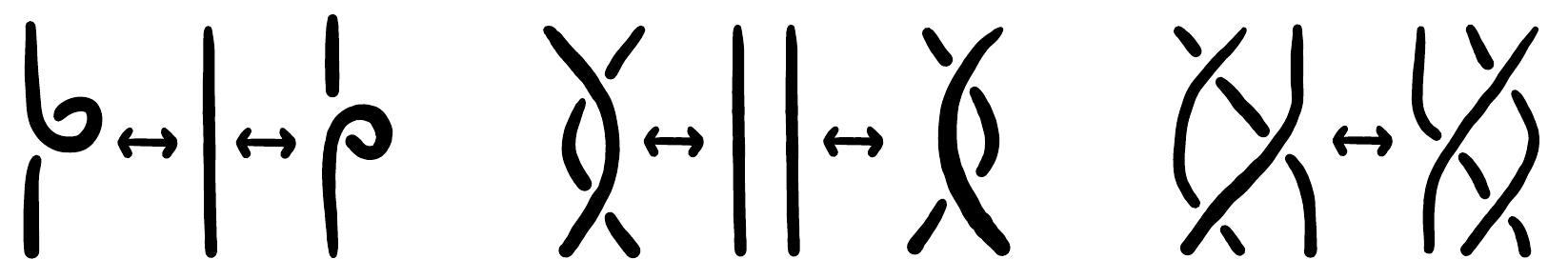}
\caption{The three types of Reidemeister moves. (We have suppressed orientations, which lead to several subcases for each type.)}
\label{fig:moves}
\end{figure}

\begin{reid}
Two oriented link diagrams $D_0$ and $D_1$ represent equivalent links $L_{D_0}$ and $L_{D_1}$ if and only if $D_0$ can be converted to $D_1$ using a sequence of the three types of local diagrammatic changes shown in Figure \ref{fig:moves} together with planar isotopies (which simple wiggle the diagrams in their plane without changing crossing information).
\end{reid}

We refer the reader to textbooks on knot theory like \cite{CrowellFox,Rolfsen} for a proof of Reidemeister's Theorem.

The local diagrammatic changes in Figure \ref{fig:moves} will be called \emph{Reidemeister moves}, and
we say two diagrams $D_0$ and $D_1$ are \emph{equivalent} if they are related by a sequence of Reidemeister moves (and planar isotopies).
With this language, Reidemeister's theorem says that two oriented link diagrams are equivalent if and only if the links they represent are equivalent.

Reidemeister's Theorem and the fact that every link has a diagram provide our starting point for the computational study of knots and links.
We will always study computational problems where the input is a link \emph{diagram}.
So from now on, in order to ease notation, we will use letters such as $L$ and $K$ to denote oriented \emph{diagrams} of links and knots, rather than smooth maps $L: \bigsqcup_{i=1}^n S^1 \to \RR^3$ and $K: \bigsqcup_{i=1}^n S^1 \to \RR^3$.

There are several different conventional ways for encoding link diagrams with a finite amount of data, such as PD codes, Gauss codes, Dowker-Thistlethwaite codes, \emph{etc}.
The specific way is not so important for the purposes of qualitative complexity theory, but it is necessary to agree on how we should measure the size of a diagram, since we will be doing things like measuring running time as a function of this size.
To this end, we define the \emph{size} of a link diagram $L$ to be
\[ |L| \defeq n+c \in \mathbb{N},\]
where $n$ is the number of crossings of $L$ and $c$ is the number of components.
For any of the standard ways of encoding a diagram suggested above, the amount of space required to store the diagram in a computer's memory is a linear function of $|L|$.

We can now precisely formulate the ``fundamental problem" of link theory as a kind of computational problem---specifically, a decision problem.
\begin{quote}
\noindent \underline{\bf Link Equivalence Problem} \\
{\bf Input:} two oriented link diagrams $L_0$ and $L_1$ \\
{\bf Output:} YES if $L_0$ and $L_1$ are equivalent diagrams; NO otherwise.
\end{quote}
Much (but by no means all) of the mathematics of knot and link theory can be motivated by the goal of finding the best possible algorithm to solve the Link Equivalence Problem.
Reidemeister's theorem implies that the Link Equivalence Problem is in the complexity class $\mathsf{RE}$ of \emph{recursively enumerable} decision problems that admit ``half" an algorithm.
Put simply, if two link diagrams represent equivalent links, then after some unbounded-but-finite amount of time (and effort), we can confirm that this is the case.
The gist of the idea is to ``brute force" search, by applying all possible sequences of Reidemeister moves to $L_0$ until we see $L_1$.
Because there is no \emph{a priori} bound guaranteed in Reidemeister's Theorem on how long this might need to take, if $L_0$ and $L_1$ are \emph{not} equivalent, then this process will never halt---it will search in vain forever.

In fact, thanks to more recent (and rather advanced) ideas involving the famous \emph{Geometrization Theorem}---conjec-tured by Thurston in 1982 \cite{Thurston-conjecture} and proved by Perelman in 2006 via Hamilton's Ricci flow method \cite{perelman2002entropy,Perelman-1,Perelman-2}---it is now generally understood that there exists an algorithm to solve the homeomorphism problem for all ``finite-volume" 3-dimensional manifolds, of which knots and links in $S^3$ are a special case; see \cite{Kuperberg-homeomorphism} for details and discussion.
Indeed, as explained in \emph{loc.\ cit.}, the Geometrization Theorem implies---as a not-too-complicated corollary---that the 3-manifold homeomorphism problem is in $\mathsf{R}$, the class of \emph{recursive} (or \emph{computable}) decision problems.
Going beyond mere \emph{computability}, Kuperberg \cite{Kuperberg-homeomorphism} furthermore shows that there exists an algorithm whose running time is bounded by a tower of exponentials, meaning the 3-manifold homeomorphism problem is in $\mathsf{ER}$, the class of \emph{elementary recursive} decision problems.

All of this is to say that there exists \emph{an} algorithm for the Link Equivalence Problem, with \emph{some} bound on the running time, but it is possible there exists a much better one.
Moreover, there are currently no known lower bounds, meaning it is entirely possible that the Link Equivalence Problem is in $\p$!
Resolving these matters is a subject of much active research by many people these days.

The Link Equivalence Problem contains several important subproblems of independent interest.
The first is simply the case where we consider \emph{knots} specifically (rather than links with an arbitrary number of components).
\begin{quote}
\noindent \underline{\bf Knot Equivalence Problem} \\
{\bf Input:} two oriented knot diagrams $K_0$ and $K_1$ \\
{\bf Output:} YES if $K_0$ and $K_1$ are equivalent diagrams; NO otherwise.
\end{quote}
To our knowledge, the current best known algorithms for the Knot Equivalence Problem are not qualitatively better than those for the more general Link Equivalence Problem.

The other important special cases of the Link Equivalence Problem are the ``recognition problems" for specific links or families of links.
The \emph{unknot} is the unique knot that is equivalent to a diagram with 0 crossings.
More generally, the \emph{$n$ component unlink} is the unique $n$ component link that is equivalent to a diagram with 0 crossings.
An \emph{unlink} is then any link equivalent to an $n$ component unlink for some $n$.
Of course, an unknot or unlink can have a diagram that does not obviously ``look like" it is an unknot or unlink;
one can consult \cite{hardunknots,hardsplitlinks} for many examples and related constructions.
It is an interesting computational problem to try to recognize unknots and unlinks.
\begin{quote}
\noindent \underline{\bf Unlink Recognition Problem} \\
{\bf Input:} an oriented link diagram $L$ \\
{\bf Output:} YES if $L$ is a diagram of an unlink; NO otherwise.
\end{quote}

\begin{quote}
\noindent \underline{\bf Unknot Recognition Problem} \\
{\bf Input:} an oriented knot diagram $K$ \\
{\bf Output:} YES if $K$ is a diagram of the unknot; NO otherwise.
\end{quote}

Both Unlink Recognition and Unknot Recognition are known to be in $\NP$.
This was first proved for Unknot Recognition in 1999 by Hass, Lagarias and Pippenger, in arguably the first ``foundational" result of complexity-theoretic knot theory \cite{HassLagariasPippenger}.
The ``certificate of unknottedness" they used was based on Haken's normal surface theory, but Lackenby has since given a new proof that uses sequences of Reidemeister moves as the certificate \cite{Lackenby-bound}.
The methods of \cite{HassLagariasPippenger} can also be used to show that Unlink Recognition is in $\NP$ (although this is not stated explicitly \emph{in loc. cit}).

Unknot Recognition is also known to be in $\mathsf{coNP}$ \cite{Kuperberg-GRH,Lackenby-norm} (but we are not aware of any such result for Unlink Recognition).
As problems in $\NP \cap \mathsf{coNP}$ are not expected to be $\NP$-hard (as this would cause a collapse of the polynomial hierarchy), this makes Unknot Recognition a reasonable candidate for admitting a polynomial time algorithm (perhaps a \emph{quantum} one?).
In fact, Lackenby has recently claimed that Unknot Recognition admits a quasi-polynomial time algorithm.

\begin{figure}
    \begin{subfigure}[t]{0.45\textwidth}
    \centering
    \includegraphics[width=\textwidth]{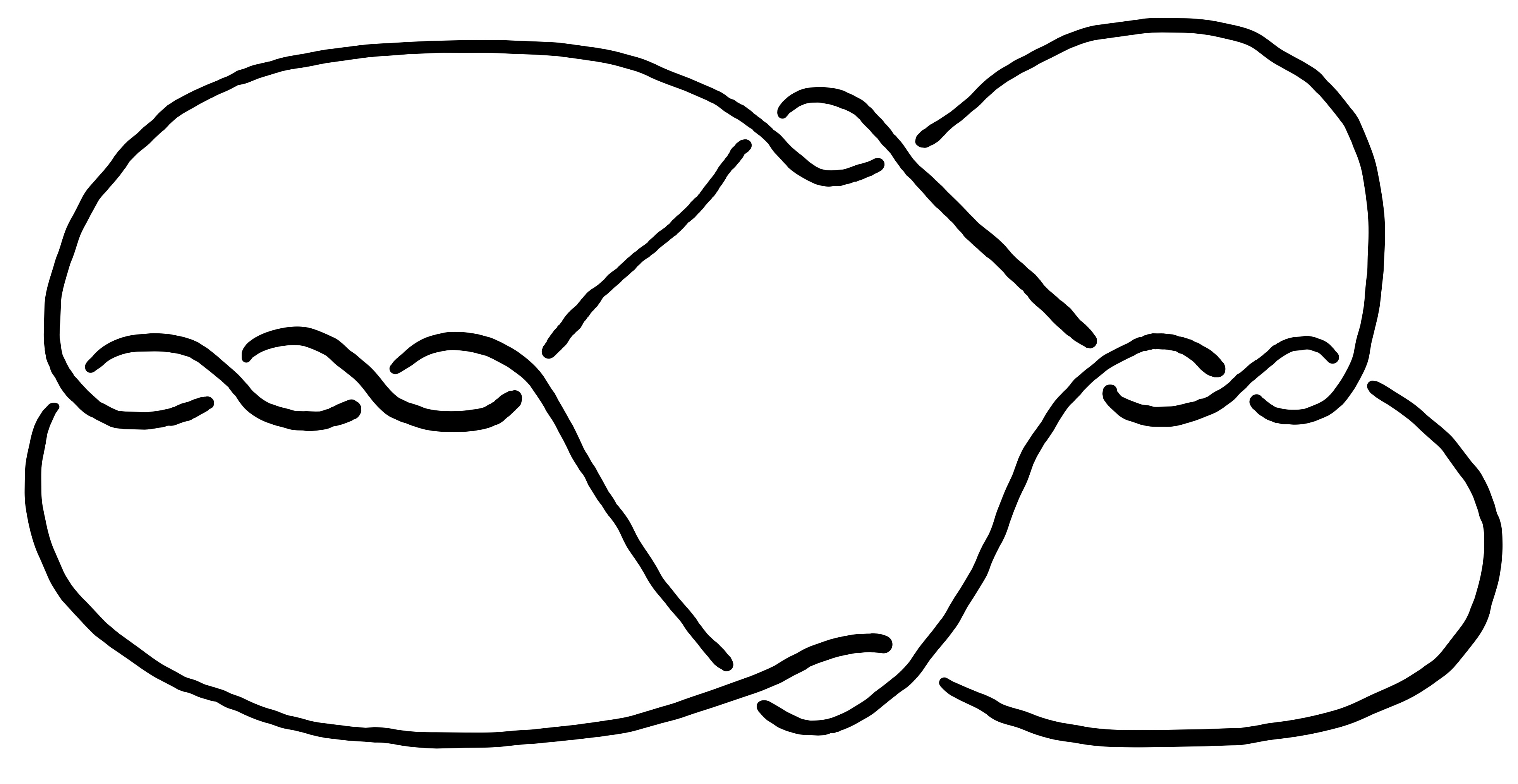}
    \caption{A hard unknot due to Goeritz \cite{Goeritz}.  Turning this diagram into the unknot via Reidemeister moves requires increasing the number of crossings.}
    \end{subfigure}
    \qquad
    \begin{subfigure}[t]{0.45\textwidth}
    \centering
    \includegraphics[width=\textwidth]{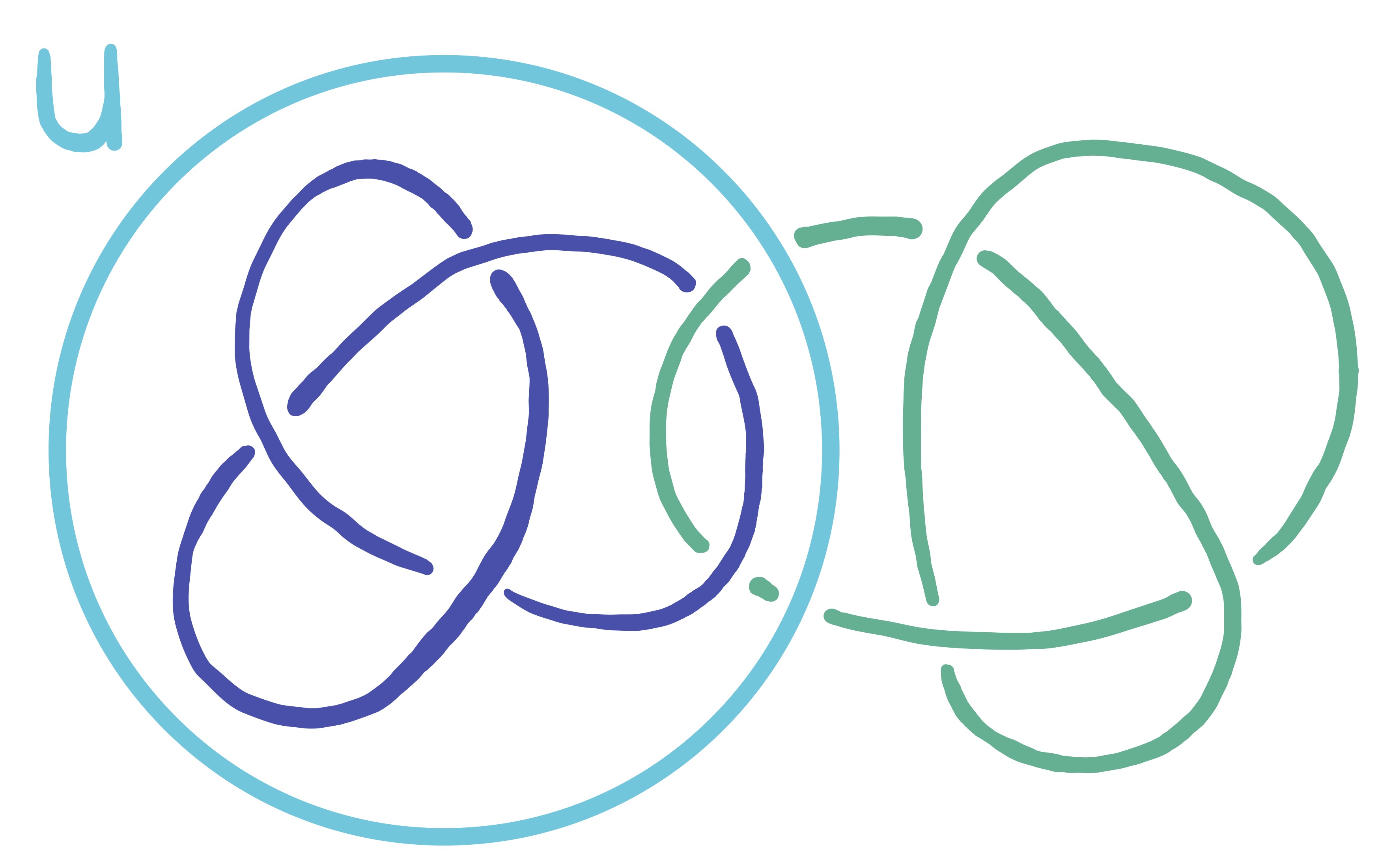}
    \caption{A hard split link due to \cite{hardsplitlinks}.  Moving the unknotted component $U$ off of the underlying two components via Reidemeister moves requires increasing the number of crossings.}
    \end{subfigure}
    \caption{}
    \label{fig-hard}
\end{figure}

One hint at the practical difficulty of solving the Unknot and Unlink Recognition problems comes from the existence of ``hard unknot" diagrams.
These are diagrams of unknots where it is \emph{proven} that there is no way to simplify them ``monotonically".  That is, for a hard unknot, if we wish to use Reidemeister moves to turn them into the standard crossing-less unknot or unlink diagram, at some point along the way, we \emph{must increase} the number of crossings before we can get rid of all of them; even worse, there exist \emph{families} of unknot diagrams where one must increase the crossings more and more, in an unbounded way.
Figure \ref{fig-hard}(A) gives an example of one of the first known hard unknots, due to Goeritz \cite{Goeritz}.
Interestingly, there are as yet no known examples of families of hard \emph{unlink} diagrams, even for two component links.
There are many sensible candidates, but actually \emph{proving} that a diagram can not be simplified monotonically is currently a difficult problem.
Figure \ref{fig-hard}(B) gives an example of a hard three component ``split link"---that is, it has one component that can be unlinked from the other two, but only after first increasing the number of crossings---first discovered in \cite{hardsplitlinks}.

For the sake of space, we will not provide any further review beyond the highlights already mentioned.
For general mathematical background on knot theory, we recommend the textbooks \cite{CrowellFox,Rolfsen,Lickorish,Adams}.
For more discussion of algorithms and complexity in knot theory, here is a list of papers (listed in chronological order, and most of which we have cited elsewhere in this document) that could serve as a starting point for more discovery:
\cite{Welsh,HassLagariasPippenger,Kuperberg-GRH,Lackenby-bound,Lackenby-norm,lackenby2017some,Koenig2021,de2021unbearable,hardunknots,hardsplitlinks,alternating}.

\subsection{Main result}
\label{ss-main}
We need two more pieces of language.
First, if $L$ is a link with $n$ components, then a \emph{sublink} of $L$ is any link we can form by forgetting about some of the components of $L$.
Second, if we say a link is \emph{trivial}, we simply mean this to be a synonym for ``is equivalent to an unlink."
We can now define our main problem of interest.

\begin{quote}
\noindent \underline{\bf Trivial Sublink Problem} \\
{\bf Input:} an oriented link diagram $L$ and a positive integer $k$ \\
{\bf Output:} YES if $L$ contains a $k$ component sublink that is trivial; NO otherwise.
\end{quote}

We note that the Trivial Sublink Problem can be formulated without requiring that $L$ be oriented.
However, orientations will be useful in our proof, so we keep them.
We also stress that it is allowed to take $k=0$ or $k=n$, where $n$ is the number of components of $L$---that is, we might forget \emph{none} of the components when we pass to a sublink, or we might forget \emph{all} of them.

Our main result is then the following.

\begin{thm}
\label{thm-main}
The Trivial Sublink Problem is $\NP$-hard.
\end{thm}

In fact, the Trivial Sublink Problem is $\NP$-complete.
This is because the methods of \cite{HassLagariasPippenger} imply (more-or-less immediately) that the Trivial Sublink Problem is in $\NP$.
However, it would take us too far afield to spend time on this, and in any case, the narrower $\NP$-hardness result of Theorem \ref{thm-main} is all we really need to accomplish our stated goal: showing in a precise sense that ``knots are complex."

As we alluded to in Subsection \ref{ss-motivations}, Theorem \ref{thm-main} has already been proved before, by two separate groups: de Mesmay, Rieck, Sedgwick and Tancer \cite{de2021unbearable}, and Koenig and Tsvietkova \cite{Koenig2021}.
Both proofs proceed by reducing from $\mathsf{3SAT}$, and involve what are called \emph{Brunnian links}: non-trivial links with the property that removing any one component always gives an unlink.
The simplest example with at least 3 components is the \emph{Borromean rings}, illustrated in Figure \ref{fig-borromean}.

In the remainder of this paper, we give a new proof of Theorem \ref{thm-main}.
Here is the idea in a nutshell:
given a graph $G$, we can build (in polynomial time) a link diagram $L_G$ such that the linking matrix of $L_G$ equals the adjacency matrix of $G$ and the trivial sublinks of $L_G$ correspond bijectively with the independent sets of $G$.
Karp famously showed the Clique Problem is $\NP$-complete \cite{karp}, from which it follows easily that the Independent Set Problem is also $\NP$-complete (see Corollary \ref{cor-ISP}), and hence the Trivial Sublink Problem is $\NP$-hard.

In the details, there are many ways one can imagine realizing such a reduction $G \mapsto L_G$.
As he explained to us in an email, Tancer originally had in mind something along the lines of what is shown in Figure \ref{fig-martin2}: build a ``general position" immersion of $G$ in the plane, and then turn each vertex $v$ into an ``octopus" (perhaps ``$\delta(v)$-pus" would be more accurate, where $\delta(v)$ is the degree of $v$) that links with each of its neighbors with a single clasp (see also \cite[\S6.2.1]{demesmay}).
This surely works, and it is supposed to be more-or-less obvious to experts that this can be achieved in polynomial time---at least in principle.
In practice, we believe we have found a reduction that is just as intuitive as Figure \ref{fig-martin2}, but is even easier to implement with an actual computer program.
The trick is to use \emph{braid words}, as we will explain in due course.

We argue that this new proof strategy via Independent Set is simpler than the proofs found in \cite{de2021unbearable} and \cite{Koenig2021}, in two senses.

First, on a more technical level, we entirely avoid the use of Brunnian links in our reduction.
The Brunnian property is quite subtle, requiring all of the components to simultaneously interact with each other in a ``global" manner.
It takes a certain amount of cleverness and background to be able to show that for every positive integer $n$, there exist $n$ component Brunnian links.
The ``gadgets" in our proof, on the other hand, only require ``pairwise" considerations between the components of the link $L_G$, mimicking the pairwise relations between vertices in $G$ described by the edges.
Informally, this can be interpreted as showing that the kind of subtle ``higher order linking" needed to analyze Brunnian links (initiated in its modern form by Milnor's theory of homotopy links \cite{Milnor}) is not a necessary ingredient for establishing hardness in link problems.
Moreover, we note for experts that while we will use integral linking numbers in order to ease our exposition, we could just as well have worked with mod 2 linking numbers; so our proof makes explicit the sense in which linking problems can be hard purely for ``mod 2 pairwise linking" reasons.

Second, on a pedagogical level, our proof involves a minimal amount of background---from both CS \emph{and} knot theory simultaneously---making it arguably the easiest approach to proving to non-experts that ``knots are complex."
To illustrate this point, building on some of SnapPy's functionality for link diagrams \cite{SnapPy} and some other Python libraries, we have created a web app visualizer that allows a user to input a graph $G$ by hand, and then see as output the link $L_G$.
It can be found by browsing to the following URL:
\begin{quote}
\centering
\url{https://trivial-sublink-git-main-shannonc8s-projects.vercel.app}
\end{quote}

We conclude our introduction by noting an additional benefit of the reduction from Independent Set:
since approximating the size of a maximum clique size is $\NP$-hard \cite{feige1991approximating,arora1998proof}, it should follow that approximating the number of components of a maximum trivial sublink is $\NP$-hard.
A similar conclusion could likely be made using the methods of \cite{de2021unbearable} or \cite{Koenig2021}, but neither discuss this point.

The next Section \ref{sec-proof} constitutes the remainder of this paper, and contains our proof of Theorem \ref{thm-main}.
However, we encourage our readers to play around with our web app before proceeding---it is possible you will be able to recover a proof of Theorem \ref{thm-main} all for yourself after doing so!

\section{Proof of Theorem \ref{thm-main}}
\label{sec-proof}

\subsection{Graphs, cliques and independent sets}
\label{ss-graphs}
Recall that a \emph{(simple, unoriented) graph} $G = (V,E)$ with vertex set $V$ is a symmetric binary relation $E \subseteq V \times V$ on $V$ that is completely irreflexive (meaning that for each $v \in V$, we have $(v,v) \notin E$).
We call the elements of $E$ the \emph{edges} of $G$.
By our assumptions, we might as well consider $E$ as a subset of the set of all \emph{unordered} pairs of \emph{distinct} elements of $V$.

If the graph has $|V|=n$ vertices and we order them $V=\{v_1,v_2,\dots,v_n\}$, then we can represent $G$ by a $n\times n$ matrix $A_G$---called the \emph{adjacency matrix} for $G$---whose $(i,j)$ entry is $1$ if $\{v_i,v_j\} \in E$, and $0$ otherwise.

\begin{figure}[h]
\centering
    \includegraphics[width=0.35\textwidth]{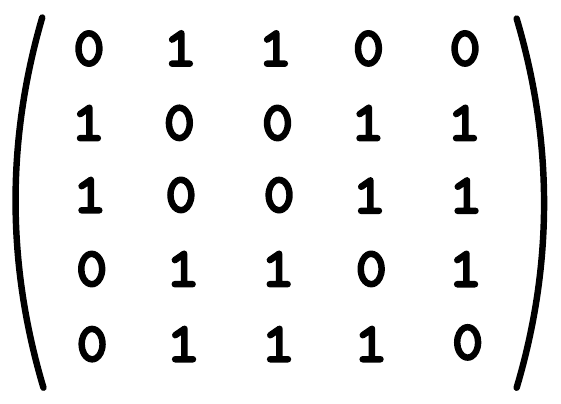}
    \caption{Adjacency matrix associated to the graph in Figure \ref{fig-martin2}.}
    \label{fig:tancers_adjacency}
\end{figure}

Our assumptions on $E$ guarantee that $A_G$ is a symmetric matrix whose diagonal entries are all 0 and whose off-diagonal entries are all either 0 or 1.
In fact, any matrix $A$ that satisfies these conditions can be understood as the adjacency matrix of a graph with $V=\{1,2,\dots,n\}$, and for computational problems whose input is a graph, we usually assume the graph is given to us in exactly such a form.

A \emph{$k$-clique} $C$ of a graph $G=(V,E)$ is a set of $k$ vertices (\emph{i.e.}, $C\subseteq V$, $|C|=k$) such that for all $v,w \in C$, $\{v,w\} \in E$.
Intuitively: if the vertices of a graph are people, and edges record friendships between those people, then a $k$-clique is a set of $k$ people who all friends with each other.
Identifying the cliques in a graph is hard, in the following sense.
\begin{quote}
\noindent \underline{\bf Clique Problem} \\
{\bf Input:} a graph $G$ (encoded by an adjacency matrix) and a positive integer $k$ \\
{\bf Output:} YES if $G$ has a $k$ clique; NO otherwise.
\end{quote}

\begin{karp}
    The Clique Problem is $\NP$-complete. \qed
\end{karp}
We note that what we are calling Karp's Theorem is really just one of the 21 different $\NP$-complete problems famously identified by Karp in 1972 \cite{karp}.
As far as the ``simplicity" of our new reduction goes, we are assuming that any reduction from one of these 21 problems is fair game.

For our purposes, it will be easier later if we consider a notion ``complementary" to the notion of clique.
A \emph{$k$-independent set} $I$ of a graph $G=(V,E)$ is a set of $k$ vertices such that for all $v,w \in I$, $\{v,w\} \notin E$.  
Building on the same intuition as earlier: a $k$-independent set is a set of $k$ people where no two of them are friends.

Cliques and independent sets are closely related.
To explain this, we use another definition:
given $G=(V,E)$, the \emph{complement graph} $G^c=(V,E^c)$ is the simple graph with the same vertex set $V$ and edges between vertices whenever $G$ does \emph{not} have such an edge.
That is: given two distinct vertices $v,w \in V$, $\{v,w\} \in E^c$ if and only if $\{v,w\} \notin E$.
It is easy to check that $(G^c)^c=G$.

Just as we formed a computational problem for cliques, we can do the same for independent sets.

\begin{quote}
\noindent \underline{\bf Independent Set Problem} \\
{\bf Input:} a graph $G$ (encoded by an adjacency matrix) and a positive integer $k$ \\
{\bf Output:} YES if $G$ has a $k$-independent set; NO otherwise.
\end{quote}

The adjacency matrix $A^c$ of $G^c$ can be constructed from the adjacency matrix $A$ of $G$ by ``flipping bits" on all of the off-diagonal entries.
That is, anywhere $A$ has a 1, we change it to a 0, and whenever an off diagonal entry of $A$ has a 0, we change it to a 1.
It is clear that we can build $A^c$ from $A$ in polynomial time in this manner.
We can formulate cliques and independent sets in terms of the adjacency matrix.
Here are the key points:
\begin{itemize}
    \item A $k$-clique $C$ of $G$ corresponds to a $k\times k$ submatrix of the adjacency matrix $A_G$ where all non-diagonal entries equal 1.
    \item A $k$-independent set $I$ of $G$ corresponds to a $k \times k$ submatrix of $A_G$ where all entries equal 0.
    \item The $k$-cliques of $G$ are exactly the same as the $k$-independent sets of $G^c$.
\end{itemize}

Combining these observations with Karp's Theorem, we arrive at a proof of

\begin{cor}
\label{cor-ISP}
    The Independent Set Problem is $\NP$-complete. \qed
\end{cor}

\subsection{Linking matrices}
\label{ss-linking}
Given a 2 component link $L$ whose components are the knots $K_1$ and $K_2$,
we now define the \emph{linking number} $\lk(K_1,K_2)$ of $K_1$ and $K_2$. It will be an element of $\ZZ$ that is a \emph{topological invariant} of $L$, in the sense that applying an ambient isotopy to $L$ will not change the value of $\lk(K_1,K_2)$.
There are many equivalent ways to formulate the definition.
We will use diagrams (as those are the most pertinent to our computational set up) and so when we say that $\lk(K_1,K_2)$ is a topological invariant of $L$, by Reidemeister's theorem, this is the same as saying that its value does not change if we apply any Reidemeister move to the diagram and compute it on the new diagram.

\begin{figure}
\centering
    \includegraphics[width=\textwidth]{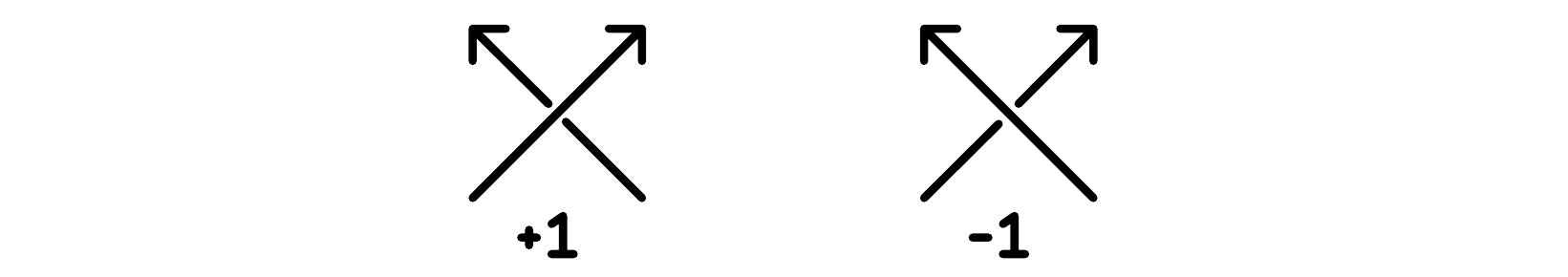}
    \caption{The two different oriented crossing types and their signs.}
    \label{fig:signs}
\end{figure}

Suppose that $L$ is given by an oriented diagram.
Up to planar isotopy (that is, up to wiggling the diagram around, but without applying a Reidemeister move), each crossing in $L$ looks like one of the two crossings shown in Figure \ref{fig:signs} (hint: you may need to tilt your head).
Given a crossing $c$ in $L$, we define $\sign(c)$ to be 1 if it looks like the crossing on the left and -1 if it looks like the one on the right.

\begin{figure}
\centering
    \includegraphics[width=0.5\textwidth]{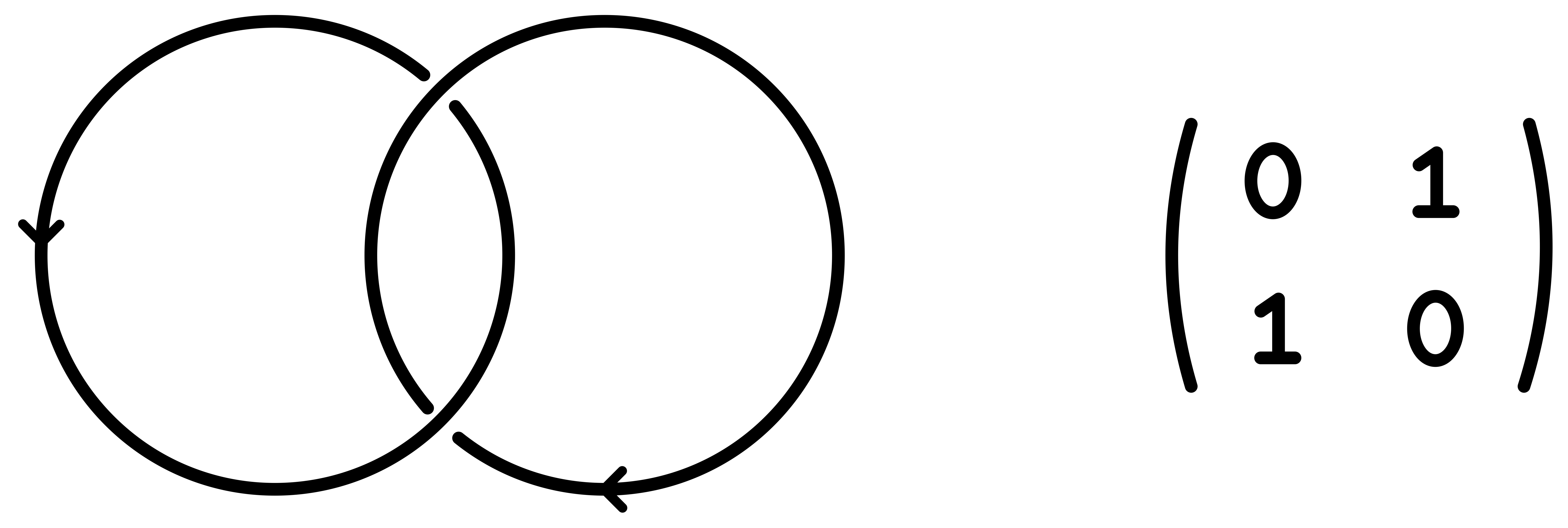}
    \caption{The Hopf link, when oriented as shown, has linking number +1.}
    \label{fig-hopf}
\end{figure}

\begin{figure}
\centering
    \includegraphics[width=14cm]{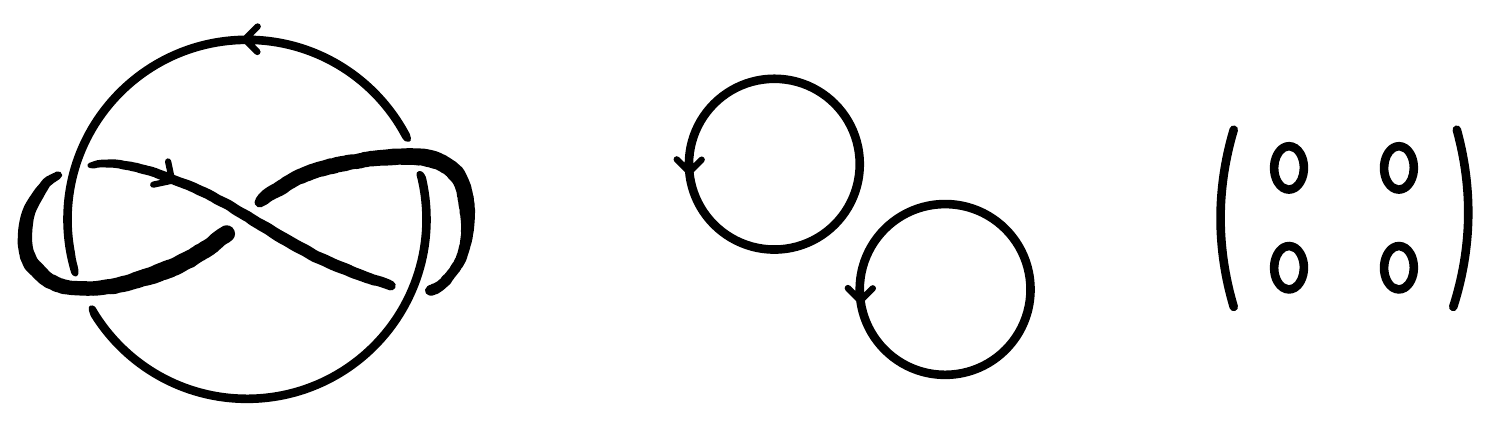}
    \caption{The Whitehead link and the 2-component unlink both have linking number 0.}
    \label{fig-whitehead}
\end{figure}

In general, $L$ will have different types of crossings according to which components are involved.
Some of the crossings will only involve the component $K_1$, some will only involve the component $K_2$ and the rest will have $K_1$ and $K_2$ crossing over and under each other.
To define $\lk(K_1,K_2)$, we sum the signs of all of the crossings of the latter type, divided by two:
\[ \lk(K_1,K_2) \defeq \sum_{\substack{c: \text{crossing involving} \\\text{ both $K_1$ and $K_2$}}} \frac{\sign(c)}{2}.\]
For example, the Hopf link shown in Figure \ref{fig-hopf} has $\lk(K_1,K_2)=1$.
The 2 component unlink and the Whitehead link shown in Figure \ref{fig-whitehead} both have $\lk(K_1,K_2)=0$.

\begin{figure}
\centering
    \includegraphics[width=.8\textwidth]{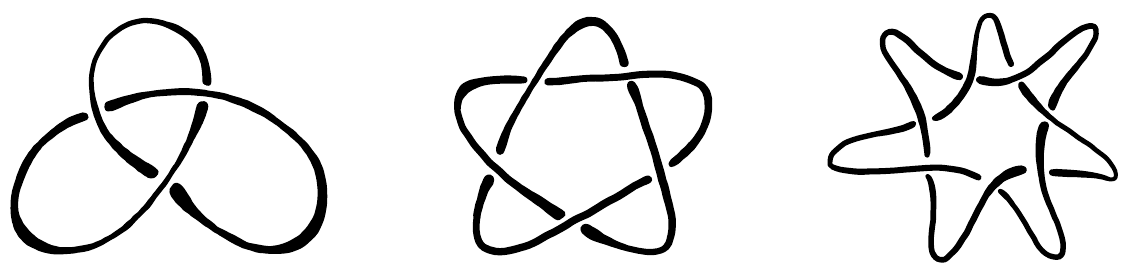}
    \caption{The $T(2,3)$, $T(2,5)$, and $T(2,7)$ torus knots.  Since they each only have one component, their linking matrices are each the $1 \times 1$ matrix with a single 0 entry.}
    \label{fig-torus_knots_linking_numbers}
\end{figure}

\begin{figure}
\centering
    \includegraphics[width=0.5\textwidth]{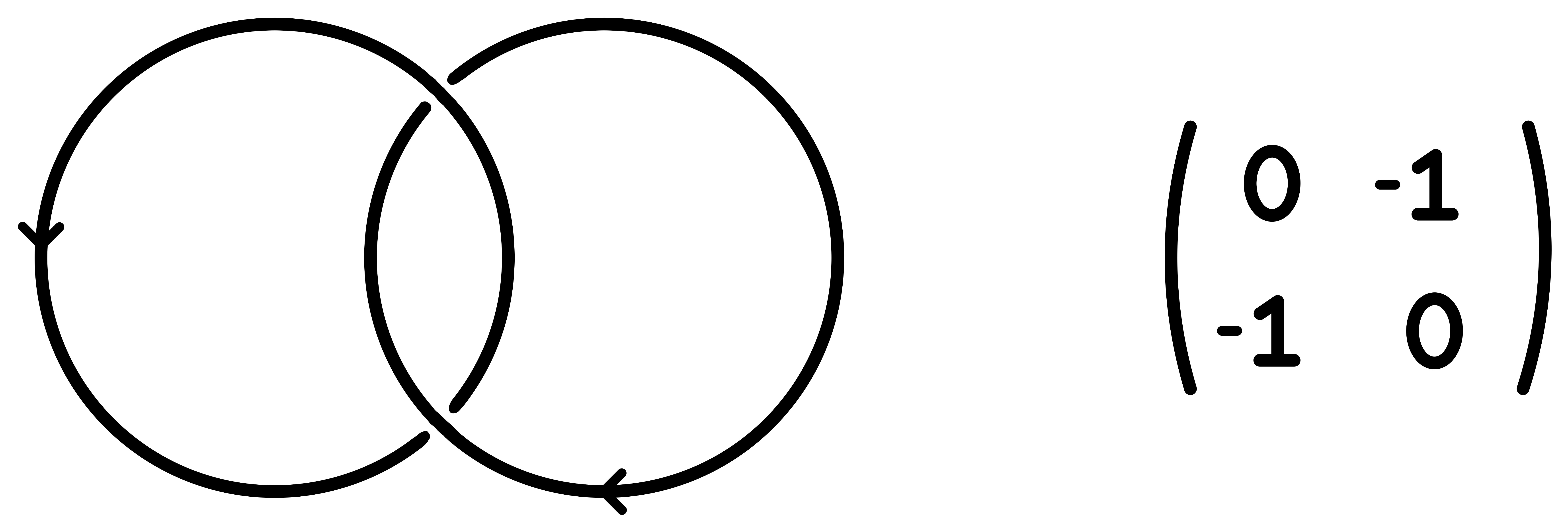}
    \caption{The Hopf link, when oriented as shown, has linking number -1.}
    \label{fig-otherhopf}
\end{figure}

\begin{figure}
\centering
    \includegraphics[width=0.5\textwidth]{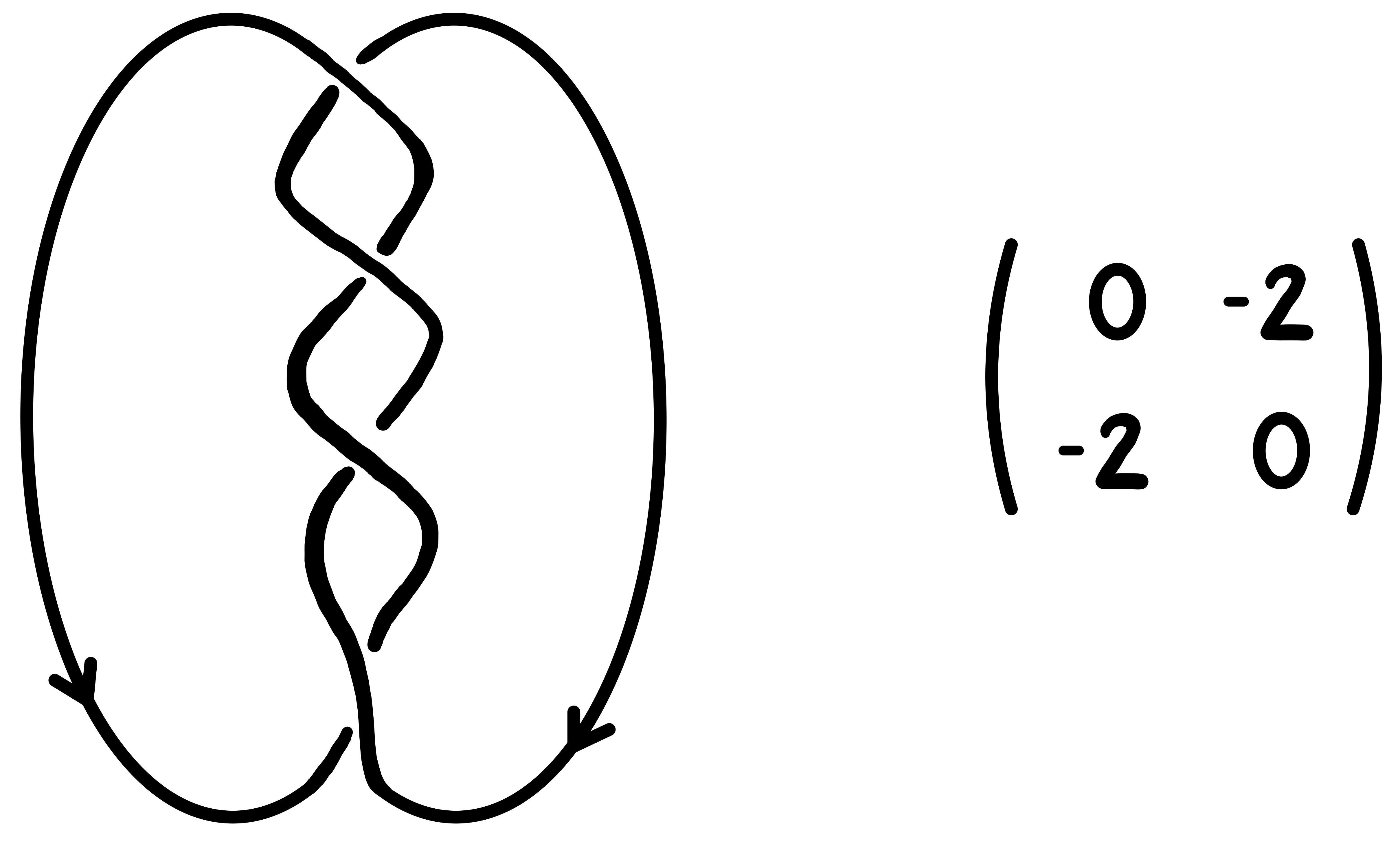}
    \caption{We can make $|lk(K_1,K_2)|$ larger by ``twisting" more.}
    \label{fig-twisting}
\end{figure}

Now we leave it as a straightforward exercise for the reader to prove

\begin{lem}
    \label{lem-lk}
    If $L$ is a 2 component link whose components are the knots $K_1$ and $K_2$, then the linking number $\lk(K_1,K_2)$ has the following properties:
    \begin{enumerate}
        \item It is always an integer: $\lk(K_1,K_2) \in \ZZ$.
        \item It is symmetric: $\lk(K_1,K_2)=\lk(K_2,K_1)$.
        \item It is a topological invariant.
        \item If $L$ is the unlink, then $\lk(K_1,K_2)=0$. \qed
    \end{enumerate}
\end{lem}

In particular, combining (2) and (3), we see that if $\lk(K_1,K_2) = 1$, then $L$ is not the 2 component unlink.

We conclude this subsection with the generalization of ``linking number" to links with more than $2$ components.
The gist is to form a square matrix whose off diagonal entries are the linking numbers of the different pairs of 2 component sublinks.
More precisely, if $L$ is an $n$ component link and we number the components $K_1,K_2,\dots,K_n$, then the \emph{linking matrix} $\lk(L)$ is the $n\times n$ matrix whose $(i,j)$ entry is $\lk(K_i,K_j)$ for $i\ne j$, and 0 otherwise.
Note that this is, by definition, a symmetric matrix, since Lemma \ref{lem-lk} implies that
\[\lk(K_i,K_j)=\lk(K_j,K_i)\]
for all $1\le i, j \le n$.

We also stress that the linking matrix makes sense even when $n=1$ or $2$.
When $n=1$, the matrix is always just the $1\times 1$ zero matrix, and when $n=2$, the matrix has zeros on its diagonal, and $\lk(K_1,K_2)=\lk(K_2,K_1)$ in its two off-diagonal entries.

We note that the linking matrix shares several similarities with the adjacency matrix.
For instance, they are always symmetric, their entries are always integers, and they always have zeroes on their diagonals.
The linking matrix also has the following properties that hark back to the properties of adjacency matrices listed at the end of the previous subsection:
\begin{itemize}
    \item The linking matrix of a sublink of $L$ is obtained by removing the rows and columns corresponding to the removed components
    \item A sublink with nonzero linking matrix is not trivial.
\end{itemize}

\begin{figure}
\centering
    \includegraphics[width=0.35\textwidth]{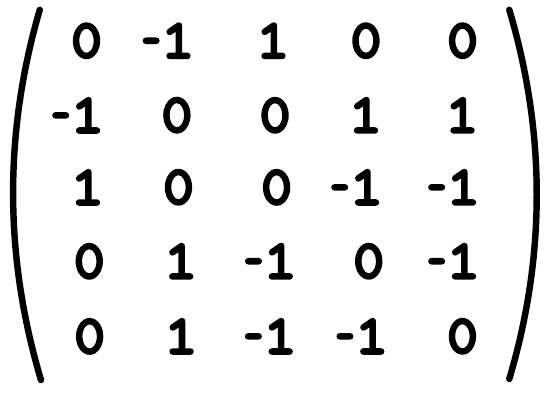}
    \caption{Linking matrix associated to the link diagram in Figure \ref{fig-martin2}.}
    \label{fig:tancers_linking}
\end{figure}

\begin{figure}
\centering
    \includegraphics[width=0.75\textwidth]{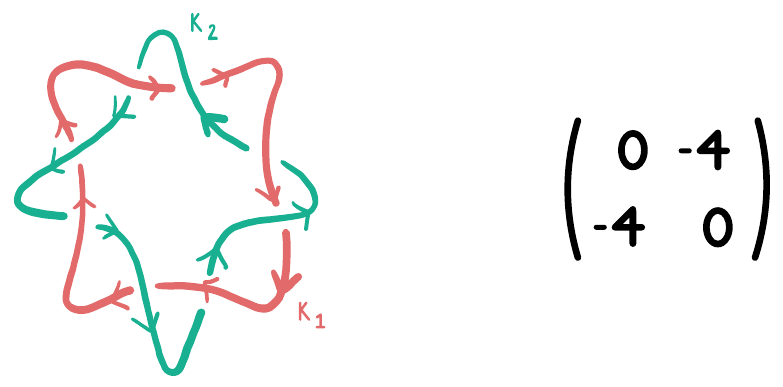}
    \caption{The $T(2,8)$ torus link and its linking matrix.}
    \label{fig:torus_knot_1}
    \end{figure}
    
\begin{figure}
\centering
    \includegraphics[width=0.75\textwidth]{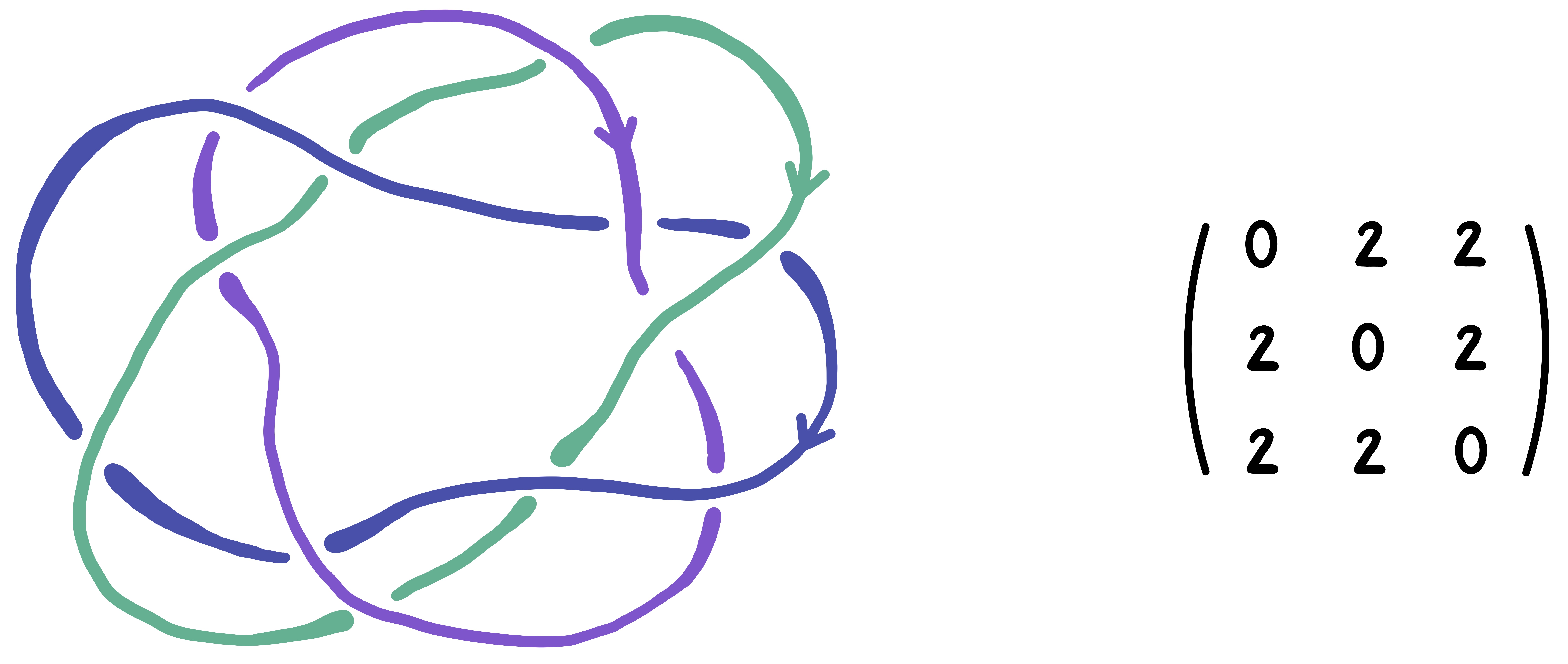}
    \caption{The $T(3,6)$ torus link and its linking matrix.}
\end{figure}

Lemma \ref{lem-lk} implies that if $L$ and $L'$ are equivalent links, then for any numbering of the components $K_1,\dots,K_n$ of $L$, there exists some numbering of the components $K_1',\dots,K_n'$ of $L'$ such that $\lk(L)=\lk(L')$.

\begin{figure}
\centering
    \includegraphics[width=0.8\textwidth]{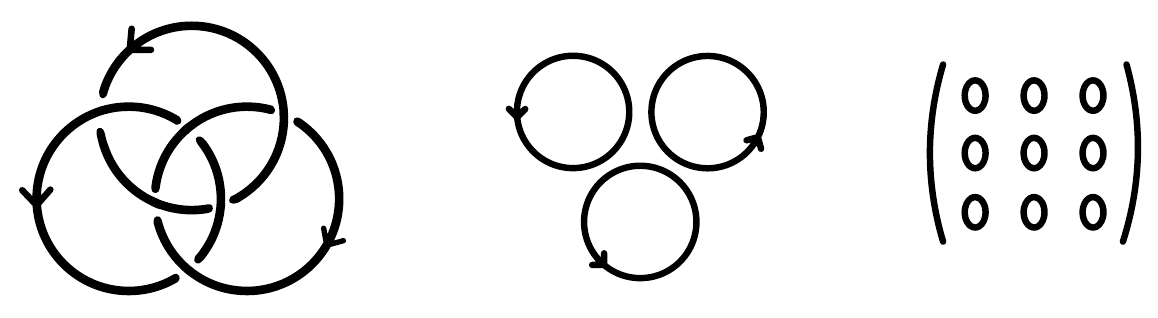}
    \caption{The Borromean rings and the 3 component unlink have the same linking matrix, consisting of all zeroes.}
\label{fig-borromean}
\end{figure}

\begin{figure}
    \centering
    \includegraphics[width=12cm]{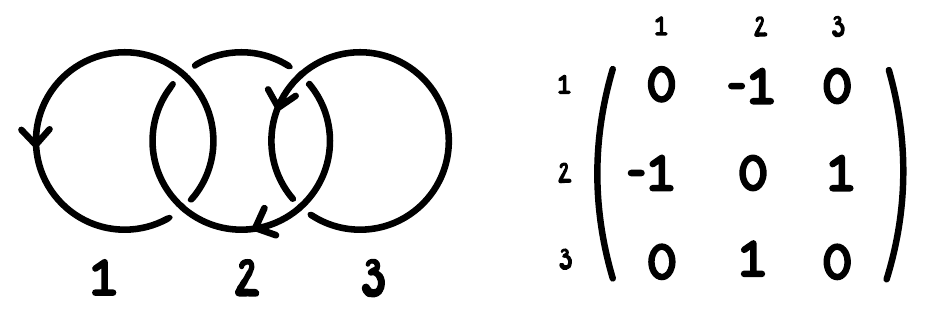}
    \caption{An example showing how we need to be careful with orientations.}
    \label{fig-openchain}
\end{figure}

Figures \ref{fig-borromean}, \ref{fig-openchain} ... give several examples of oriented links and their linking matrices.

\subsection{Braid words and their trace closures}
\label{ss-braids}
\begin{figure}
    \centering
    \includegraphics[width=0.5\linewidth]{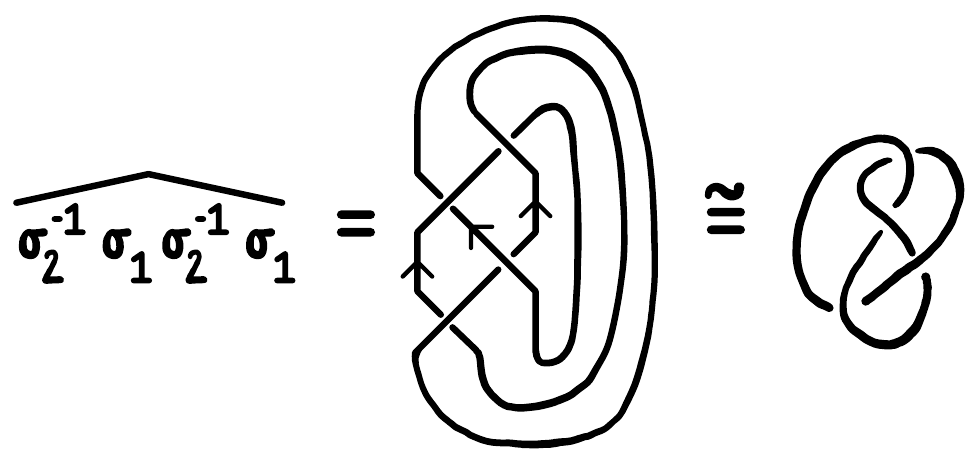}
    \caption{An example realizing the figure 8 knot as the trace closure of a braid word.}
    \label{fig-eight}
\end{figure}

An especially convenient way to generate diagrams of links is to ``close off braids."
We will not attempt to discuss this in any level of generality, nor will we make any attempt at introducing the reader to the wonderfully rich algebra of braids and the braid groups.
Before diving in, we direct the reader to the example provided in Figure \ref{fig-eight}.

Fix a positive integer $n$ that we will call the \emph{number of strands}.
The \emph{generators} of the $n$ strand braid group are then, by definition, the $n-1$ symbols $\sigma_1,\sigma_2,\dots,\sigma_{n-1}$.
A \emph{braid word on $n$ strands} is a formal product of the $\sigma_i=\sigma_i^{+1}$ and their ``inverses," \emph{i.e.}\ the symbols $\sigma_1^{-1},\sigma_2^{-1}\dots,\sigma_{n-1}^{-1}$.
For example, if $n=3$, then $w=\sigma_2^{-1}\sigma_1\sigma_2^{-1}\sigma_1$ is a braid word on 3 strands.
(Note that one needs to specify $n=3$ in order to be sure that $w$ does not happen to be a braid word on, say, 5 strands that just happens not to use $\sigma_3^{\pm 1}$ or $\sigma_4^{\pm 1}$.)

\begin{figure}
    \centering
    \includegraphics[width=15cm]{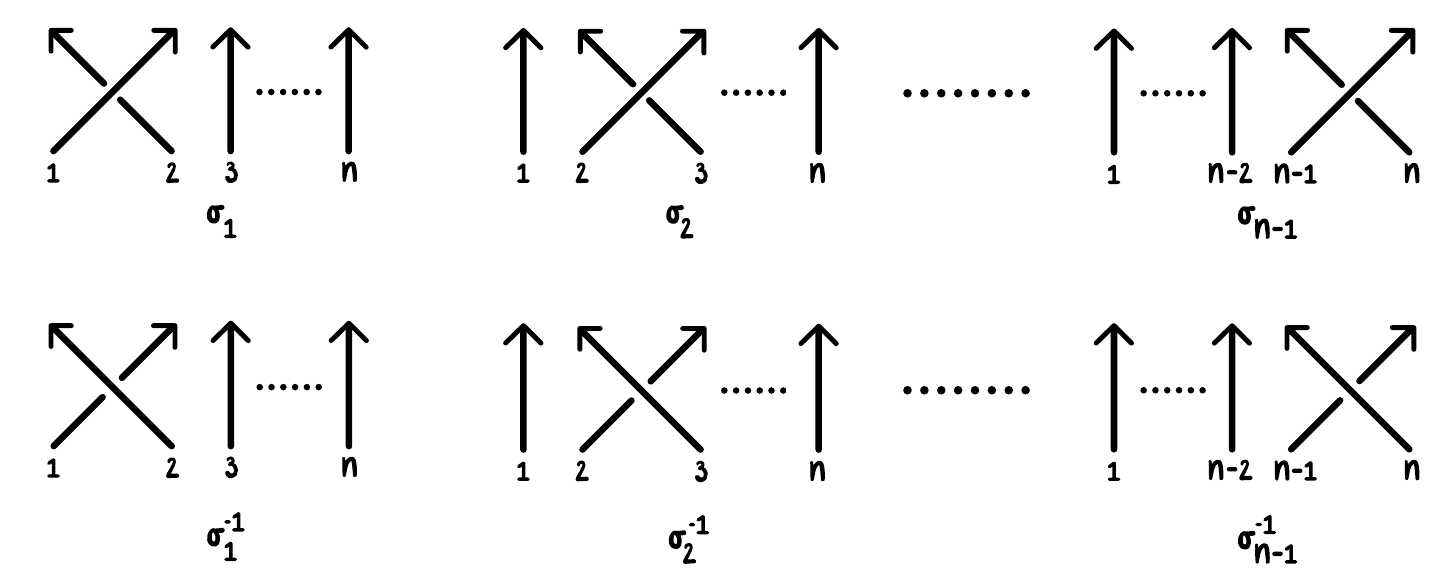} 
    \caption{The braid generators on $n$ strands, and their inverses.}
    \label{fig-gens}
\end{figure}

A braid word on $n$ strands is best understood as a combinatorial representation of a \emph{braid diagram}, which is built using two rules.
First, each generator (or inverse generator) corresponds to one of the elementary diagrams shown in Figure \ref{fig-gens}.
Second, concatenating two braid words $w_1w_2$ amounts to stacking the picture for $w_1$ on top of $w_2$.

\begin{figure}
    \centering
    \includegraphics[width=12cm]{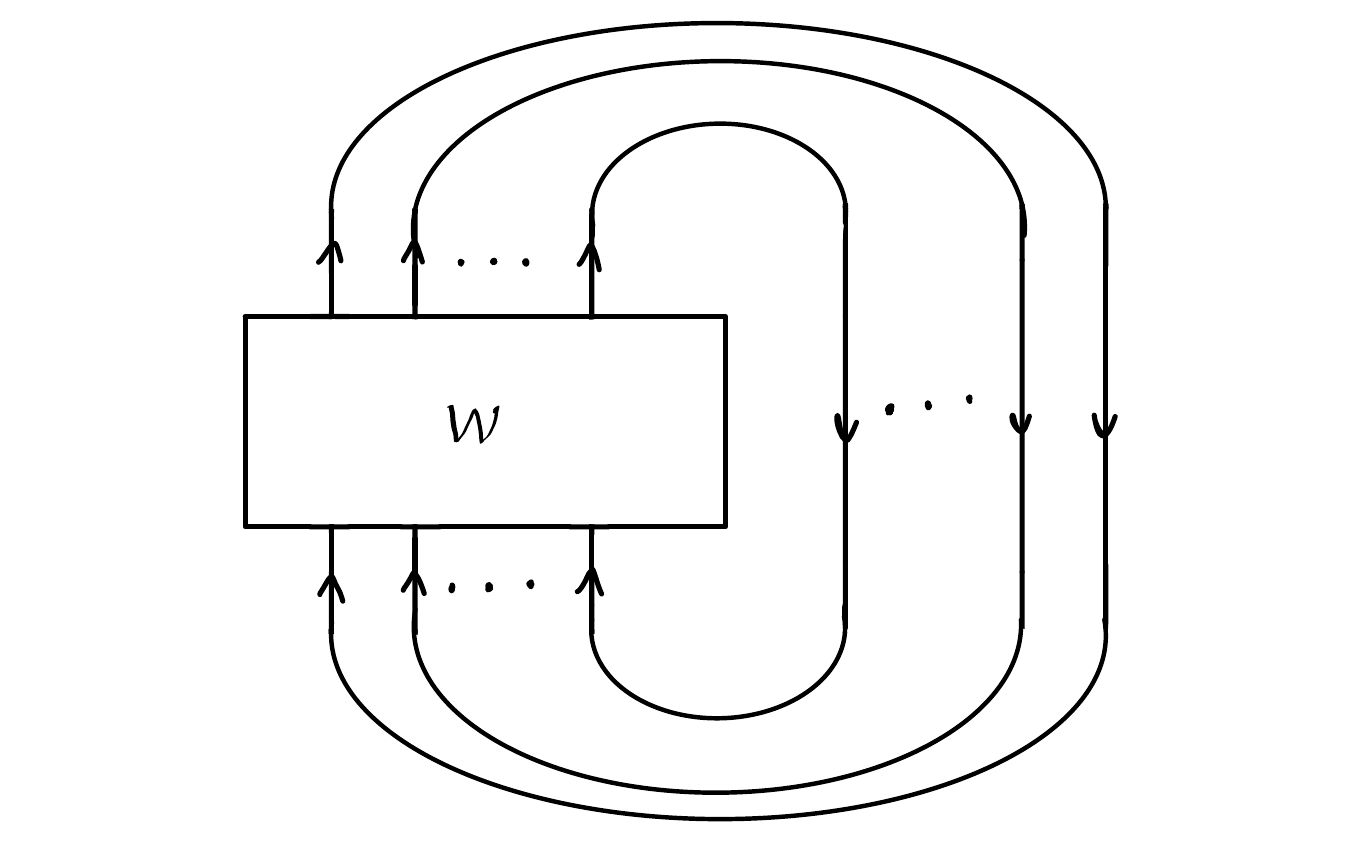}
    \caption{The \emph{trace closure} $\hat{w}$ of a general braid word $w$.}
\label{fig-trace}
\end{figure}

Given a braid word $w$, the \emph{trace closure} $\hat{w}$ is the oriented link diagram we get by connecting the top of the first strand to the bottom of the first strand, the top of the second strand to the bottom of the second strand, \emph{etc.}, in a way that creates no new crossings.
Orienting the braid strands upward induces the orientation on $\hat{w}$.
See Figure \ref{fig-trace}.

For whatever precise way we decide to encode our link diagrams (PD codes, Gauss codes, \emph{etc.}), a given braid word $w$ of length $l$ (that is, $w$ is a concatenation of $l$ of the generators and their inverses) can be turned into the link diagram $\hat{w}$ with an algorithm that has polynomial running time as a function of $l$.

\subsection{Pure braids}
\label{ss-pure}
\begin{figure}
    \centering
    \includegraphics[width=.5\textwidth]{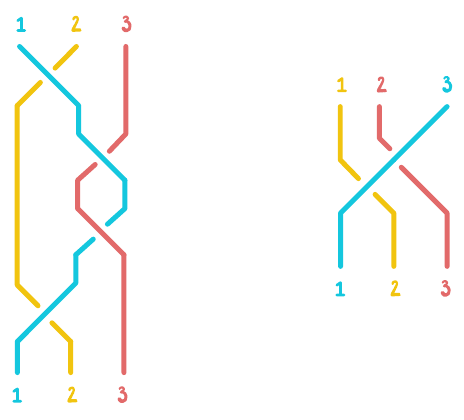}
    \caption{Example of a pure and non-pure braid, respectively.}
    \label{fig-pure-nonpure}
\end{figure}

Consider a braid word $w$ on $n$ strands. Label the strands along the bottom of the braid diagram corresponding to $w$ from left to right using $1,\dots, n$. Each strand can be traced through to the top of the braid diagram, determining a permutation $\sigma_w\in S_n$ of the $n$ strands. If strand $i$ returns to ``slot'' $i$, then $\sigma_w(i)=i$. If strand $i$ instead ends in ``slot'' $j$, then $\sigma_w(i)=j$. Notice that in Figure \ref{fig-pure-nonpure}, the braid diagrams correspond to the identity permutation $\text{id}$ and $(132)$ in $S_3$, respectively. A braid is \emph{pure} if its corresponding permutation is the identity permutation. Otherwise, it is \emph{non-pure}. By construction, the trace closure of a pure braid with $n$ strands must have $n$ components, each of which is an unknot.  More generally, the number of components of $\hat{w}$ is exactly the number of orbits of the action of $\langle \sigma_w \rangle \le S_n$ on the set of strands $\{1,2,\dots,n\}$.

\begin{figure}
    \centering
    \includegraphics[width=.5\textwidth]{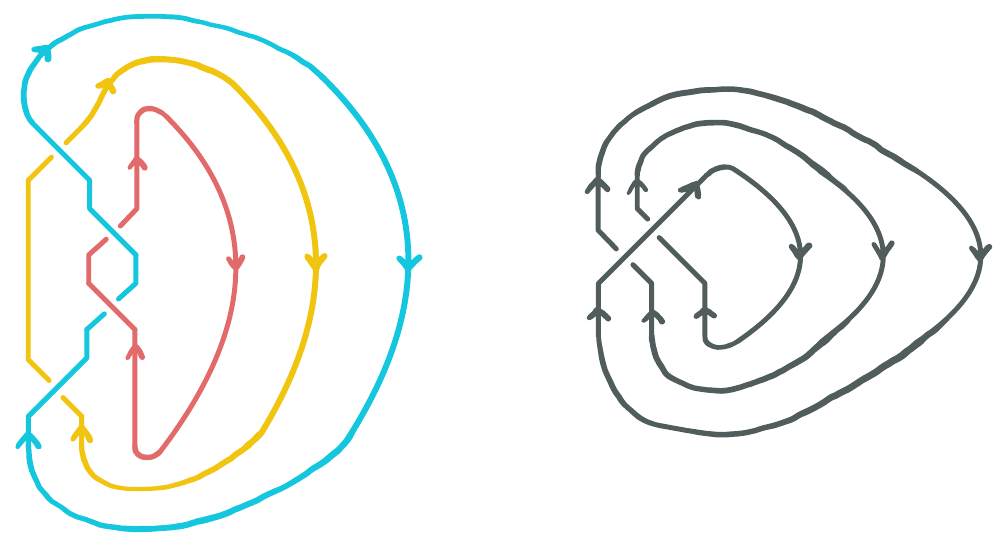}
    \caption{Trace closures corresponding to the braid diagrams in figure \ref{fig-pure-nonpure}, respectively.}
    \label{fig-trace_closures_two_braids}
\end{figure}

\subsection{An elementary splitting criterion}
\label{ss-splitting}
We need one more tool before we are in a position to reduce the Independent Set Problem to the Trivial Sublink Problem.
Intuitively, a link is called ``split" if there is a way to separate two nonempty subsets of its components into two ``unlinked" sublinks.
More precisely, a link is \emph{split} if there exists a diagram of the link that is a disjoint union of two non-empty diagrams.
Here the ``disjoint union" of two link diagrams $L_1$ and $L_2$ is the diagram formed by putting $L_1$ next to $L_2$ without introducing any crossings between $L_1$ and $L_2$.

\begin{figure}
\centering
    \includegraphics[width=.7\textwidth]{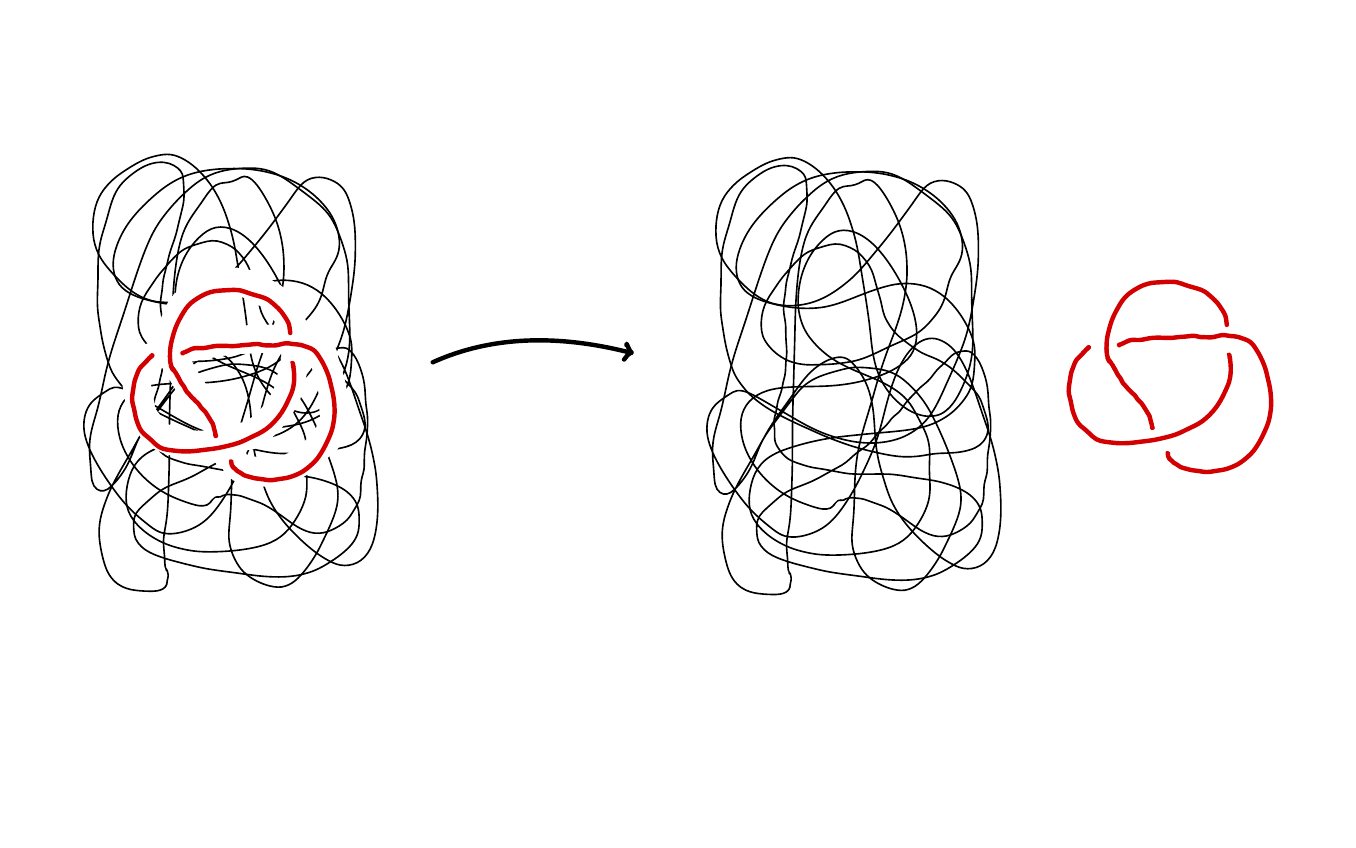}
    \vspace{-2cm} 
    \caption{No matter what ``rats nest" lies underneath (shown in black), if the red knot lies entirely on top, then we can split it off with an ambient isotopy.}
    \label{fig-nest}
\end{figure}

The final tool we need is a lemma that formalizes a very simple idea: if a link diagram has a component that sits ``on top of" all of the other components, then that component can be split off.
Figure \ref{fig-nest} illustrates what we mean.

\begin{lem}
\label{lem:ontop}
 Suppose $L=L_1 \cup \cdots \cup L_n$ is an $n$ component link with the following property: for each $k=1,\dots,n-1$, $L_n$ only ever crosses \emph{over} component $L_k$ (that is, for each crossing between $L_n$ and $L_k$, $L_n$ is always the component that passes over).  Then $L_n$ can be split from the first $n-1$ components.  In particular, $L$ is equivalent to a link where $L_n$ does not cross with any of the other components at all.
\end{lem}

We leave the proof of this lemma as an elementary-but-tedious exercise in understanding the definitions of ``split link", ``equivalence of links", and ``link diagram."
Note that while it is in principle possible to give an ``algorithmic" proof of the lemma via Reidemeister moves, it is actually quite difficult to work out all of the details!
This is a situation where using the abstract tools of topology---specifically, isotopies---is much easier for proof writing than the combinatorial tool of Reidemeister moves.

\subsection{The final reduction}
\label{ss-reduction}
We now reduce the Independent Set Problem to the Trivial Sublink Problem.
It suffices to give a polynomial time algorithm that converts an adjacency matrix $A$ for a graph on $n$ vertices into a diagram of a $n$ component link $L_A$ such that, for each $k=1,\dots,n$, the $k$ component trivial sublinks of $L_A$ are in bijection with the $k$-independent sets of the graph encoded by $A$.
Indeed, if we can show this, then for any input $(A,k)$ to the Independent Set Problem, $(L_A,k)$ will be an input to the Trivial Sublink Problem with the answer YES if and only if the answer to $(A,k)$ is YES.
The $\Rightarrow$ direction here constitues the soundness of the reduction, and $\Leftarrow$ the completeness (refer back to \S\ref{ss-complexity} for an explanation of these terms if needed).

To build $L_A$, we first use $A$ to build a particular braid word $w_A$ on $n$ strands as follows.
For each $i=1,\dots,n-1$, define
\[ w_i \defeq \sigma_i\sigma_{i+1}\cdots \sigma_{n-1}\sigma_{n-1}^{\epsilon_{i,n}}\cdots\sigma_{i+1}^{\epsilon_{i,i+2}}\sigma_i^{\epsilon_{i,i+1}}\]
where
\[ \epsilon_{i,j} \defeq \begin{cases} -1 & \text{ if } A_{i,j}=0, \\ +1 & \text{ if } A_{i,j} = 1. \end{cases}\]
Let
\[ w_A \defeq w_1w_2\cdots w_{n-1}.\]
Finally, let $L_A = \widehat{w_A}$ be the trace closure of $w_A$.

Notice that $w_A$ is a pure braid.
As explained in Subsection \ref{ss-pure}, the means each component of $L_A = \widehat{w_A}$ is an unknot, and there is exactly one component for each strand of the braid $w_A$.
So $L_A$ has $n$ components---one for each vertex of the graph $A$.
Better yet, the natural numbering of the strands of $w_A$ from left to right gives us an identification of the components $K_1,K_2,\dots,K_n$ of $L_A$ with the vertices $1,2,\dots,n$ of $A$.

\begin{figure}
\centering
\begin{tikzpicture}[]
\node (input) at (-7,-3) {\includegraphics[width=0.35\textwidth]{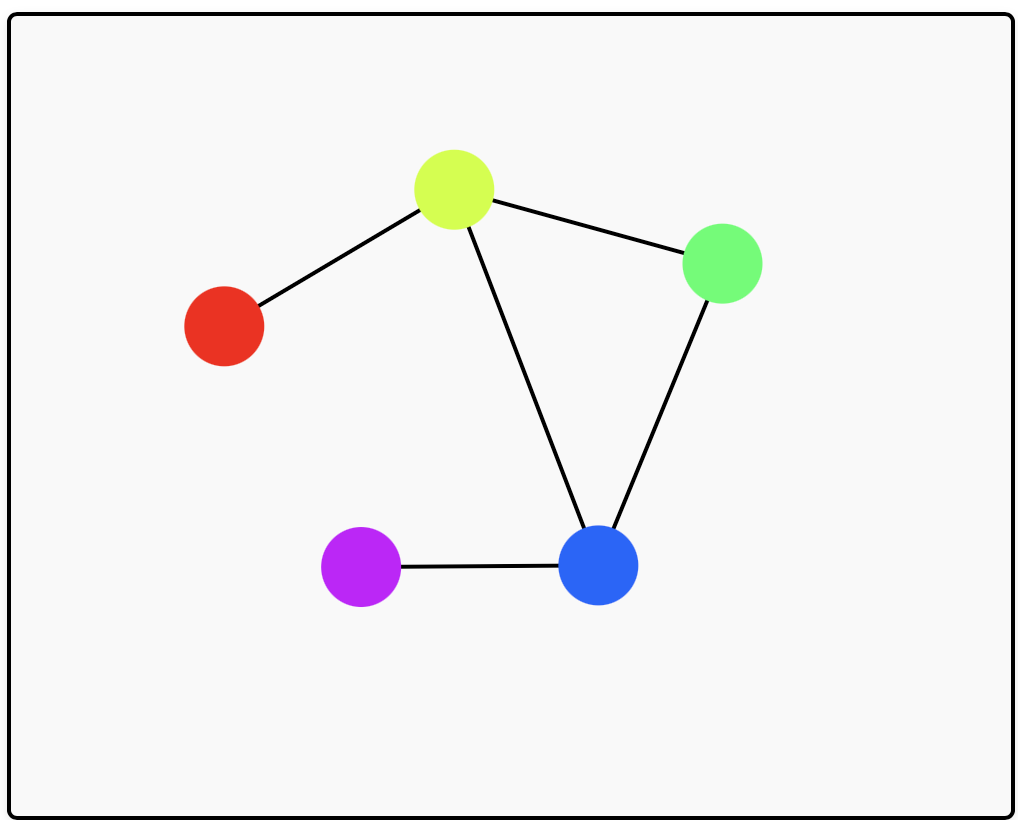}};
\draw[|->,thick] (input.south east) -- (-1,-8);
\draw (-.75,-8.25) node{$w_A$};
\pic[
    line width=1.2pt,
    braid/.cd,
    strand 1/.style={red},
    strand 2/.style={yellow},
    strand 3/.style={green},
    strand 4/.style={blue},
    strand 5/.style={purple}
]
{braid=
s_1^{-1} s_2^{-1} s_3^{-1} s_4^{-1} s_4 s_3 s_2 s_1^{-1}
s_2^{-1} s_3^{-1} s_4^{-1} s_4 s_3^{-1} s_2^{-1}
s_3^{-1} s_4^{-1} s_4 s_3^{-1}
s_4^{-1} s_4^{-1}};
\draw[red,very thick,->] (0,-12) -- (0,-11);
\draw[yellow,very thick,->] (0,-5) -- (0,-4);
\draw[green,very thick,->] (1,-12) -- (1,-11);
\draw[blue,very thick,->] (2,-12) -- (2,-11);
\draw[purple,very thick,->] (4,-8) -- (4,-7);
\draw (0,0) node[above,red] {$1$};
\draw (1,0) node[above,yellow] {$2$};
\draw (2,0) node[above,green] {$3$};
\draw (3,0) node[above,blue] {$4$};
\draw (4,0) node[above,purple] {$5$};

\draw (-7,-7) node {A=$\begin{pmatrix} 0 & 1 & 0 & 0 & 0 \\ 1 & 0 & 1 & 1 & 0 \\  0 & 1 & 0 & 1 & 0\\  0 & 1 & 1 & 0 & 1\\  0 & 0 & 0 & 1 & 0\\ \end{pmatrix}$};

\node (output) at (-6,-17) {\includegraphics[width=0.3\textwidth]{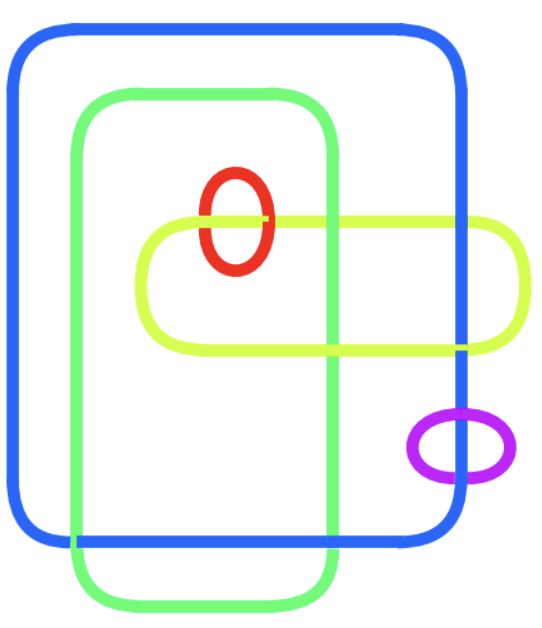}};
\draw[<-|,thick] (output.north east)-- (-1,-12);
\draw (-3.75,-14.25) node {$L_A$};
\end{tikzpicture}

\caption{An example of our reduction $A \mapsto w_A \mapsto L_A = \widehat{w_A}$, as generated by our interactive web app (which greedily eliminates extraneous crossings in $L_A$).}
\label{fig-example}
\end{figure}

An example of this construction is shown in Figure \ref{fig-example}.

Clearly the reduction $A \mapsto L_A$ can be carried out in polynomial time.
It remains to show that, for each $k$, the $k$-component trivial sublinks of $L_A$ are in bijection with the $k$-independent sets of $A$.
We already observed that each individual component is an unknot (as $w_A$ is a pure braid).
But note now that this conforms with the fact that every singleton set $C=\{v\} \subseteq V$ of a simple graph is independent.

Next, we observe that, by construction,
\[ \lk(L_A) = A.\]
That is, two components of $L_A$ have linking number 1 whenever the two components come from vertices of $A$ that share an edge, and otherwise have linking number 0.
It takes a little bit of work to check this, unraveling our definitions, so we do so now.

First note that any two strands cross exactly twice. More specifically, if we assume $m < p$ (which we may as well), then we see that the only two crossings between strand $m$ and strand $p$ show up due to the braid word $w_m$, specifically the generators $\sigma_{p-1}$ and $\sigma_{p-1}^{\epsilon_{m,p}}$ in that word.
To check this, one simply needs to draw a picture of the braid word $w_m$ and label the strands corresponding to $m$ and $p$.
By definition of $\epsilon_{m,p}$, this guarantees that for any two components, there exists exactly one oriented crossing with sign $+1$ and one other crossing with sign dependent on $A_{m, p}$.
Therefore, 
\begin{equation}
lk(K_m,K_p) = 
    \begin{cases}
        \frac{+1-1}{2} = 0 & \text{ if } A_{m,p}=0 \\
        \frac{+1+1}{2} = 1 & \text{ if } A_{m,p}=1.
    \end{cases}
\end{equation}

From these observations, we can quickly establish soundness of our reduction: suppose $L'$ is a $k$-component trivial sublink of $L$ with components $K_{i_1},\dots,K_{i_k}$.
Then
\[ A_{i,j} = \lk(K_{i_a},K_{i_b})=0 \]
for each $i_a,i_b \in \{i_1,\dots,i_k\}$.
In other words, $\{i_1,\dots,i_k\}$ is a $k$-independent set of $A$.

Finally, we establish completeness.
Suppose $\{i_1,\dots,i_k\}$ is a $k$-independent set of $A$.
Without loss of generality, we assume that $i_1 < i_2 < \cdots < i_k$.
We shall show that the sublink $L'$ of $L_A$ with components $K_{i_1},\dots,K_{i_k}$ is a trivial sublink.
Since $\lk(L_A) = A$ and $\{i_1,\dots,i_k\}$ is an independent set, it follows that each of the linking numbers between the pairs of components of $L'$ is 0.
Unfortunately, \emph{a priori}, this is not enough to guarantee that $L'$ is an unlink with $k$ components;
for example, the Borromean rings and Whitehead link show that it is possible for a linking matrix to vanish and yet the link need not be trivial.

Fortunately, the simple definition of $w_A$ allows us to apply Lemma \ref{lem:ontop}.
By construction of $w_A$ (especially the definition of the $\epsilon_{i,j}$), the fact that $\{i_1,\dots,i_k\}$ is an independent set with guarantees that $K_{i_k}$ never passes \emph{under} $K_{i_{k-1}}, \dots, K_{i_2}$, or $K_{i_1}$.
Indeed, by the definition of $w_A$, we can deduce there are only two parts of the braid word where the component $K_{i_{k}}$ might be made to pass under another component $K_j$: in $w_{i_{k}}$ if $j>i_k$ or or in $w_{j}$ if $j<i_k$. 
Since $i_1<i_2<\dots<i_k$, we deduce that any crossing of $K_{i_k}$ under $K_{i_j}$ for some $1\le j< k$ must occur due to the latter case, namely, $w_{i_j}$.
We can rule this out because  $i_j$ and $i_k$ are in an independent set, meaning $A_{{i_k},{i_j}} =A_{{i_j},{i_k}} =0$, and thus $\epsilon_{{i_k},{i_j}}=-1$.
Therefore, by Lemma \ref{lem:ontop}, $K_{i_k}$ can be split off of all the other components $K_{i_{k-1}},\dots,K_{i_1}$.
Proceeding inductively, we can split off $K_{i_{k-1}}$ from the remaining links, and so on, until each of the components $K_{i_k},\dots,K_{i_1}$ has been split off from each of the others.
Since each individual component is an unknot, we conclude that $L'$ is a trivial sublink.
\qed

\newcommand{\etalchar}[1]{$^{#1}$}

\end{document}